\documentclass[preprint]{elsarticle}
\usepackage[numbers]{natbib}
\usepackage{caption}
\usepackage{subcaption}
\usepackage{adjustbox}
\usepackage{tcolorbox}
\usepackage{booktabs}
\usepackage{multicol}
\usepackage{float}
\usepackage{multirow}
\usepackage{booktabs}
\usepackage{flushend}
\usepackage{rotating}
\newcommand{\specialcell}[2][c]{\begin{tabular}[#1]{@{}c@{}}#2\end{tabular}}
\usepackage[para,online,flushleft]{threeparttable}
\usepackage{framed}

\usepackage{framed}
\usepackage{xcolor}
\usepackage{lipsum}% dummy text
\usepackage{hyperref}

\usepackage{booktabs, multirow} % for borders and merged ranges
\usepackage{soul}% for underlines
\usepackage{changepage,threeparttable} % for wide tables

\newlength{\leftbarwidth}
\newlength{\leftbarsep}
\colorlet{leftbarcolor}{black}
\renewenvironment{leftbar}{%
    \MakeFramed {\advance \hsize -\width \FrameRestore }%
}{%
    \endMakeFramed
}

\usepackage{mwe}

\usepackage{amsmath,amssymb,amsfonts}
\usepackage{algorithmic}
\usepackage{xcolor}
\def\BibTeX{{\rm B\kern-.05em{\sc i\kern-.025em b}\kern-.08em
    T\kern-.1667em\lower.7ex\hbox{E}\kern-.125emX}}
    
\newcommand{\heng}[1]{\textcolor{red}{{\it [Heng says: #1]}}}
\newcommand{\bhagya}[1]{\textcolor{green}{{\it [Bhagya says: #1]}}}
\newcommand{\review}[2]{\textcolor{black}{{#2}}}

\begin{document}

%\author{Bhagya Chembakottu\\Polytehcnique Montreal \And Heng Li\\Polytechnique Montreal\And Foutse Khomh\\Polytechnique Montreal}
% \Plainauthor{Achim Zeileis, Second Author}

%\title{A Large-Scale Empirical Study of Android Sports Apps in the \proglang{Google Play Store}}

%\Plaintitle{A Large-Scale Empirical Study of Android Sports Apps in the Google Play Store}
%\Shorttitle{Study of Android Sports Apps in the \proglang{Google Play Store}}

\title{A Large-Scale Exploratory Study of Android Sports Apps in the Google Play Store}

%\thanks{AppBrain: https://www.appbrain.com/, 42matters: https://42matters.com/}
%}

\author{Bhagya Chembakottu}
\ead{bhagya.c@polymtl.ca}
\author{Heng Li}
\ead{heng.li@polymtl.ca}
\author{Foutse Khomh}
\ead{foutse.khomh@polymtl.ca}
\address{Dept. of Computer and Software Engineering\\Polytechnique Montreal, Canada}

%\author{\IEEEauthorblockN{Bhagya Chembakottu, Heng Li, Foutse Khomh}\\
%\IEEEauthorblockA{\textit{Dept. of Computer and Software Engineering} \\
%\textit{Polytechnique Montreal}\\
%bhagya.c@polymtl.ca, heng.li@polymtl.ca, foutse.khomh@polymtl.ca}
%}

%% - \Abstract{} almost as usual
\begin{abstract}
 \noindent \textbf{Context:} Prior studies on mobile app analysis often analyze apps across different categories or focus on a small set of apps within a category. These studies either provide general insights for an entire app store which consists of millions of apps, or provide specific insights for a small set of apps. However, a single app category can often contain tens of thousands to hundreds of thousands of apps. For example, according to AppBrain, there are 46,625 apps in the ``Sports'' category of Google Play apps. Analyzing such a targeted category of apps can provide more specific insights than analyzing apps across categories while still benefiting many app developers interested in the category.
  
 \noindent \textbf{Objective:} This work aims to study a large number of apps from a single category (i.e., the \textit{sports} category). Our work can provide two folds contributions: 1) identifying insights that are specific to tens of thousands of sports apps, and 2) providing empirical evidence on the benefits of analyzing apps in a specific category. %; and 3) providing lessons learned from sports apps development that can benefit general app development. 
  
 \noindent\textbf{Method:} We perform an empirical study on over two thousand sports apps in the Google Play Store. 
  We study the characteristics of these apps (e.g., their targeted sports types and main functionalities) through manual analysis, the topics in the user review through topic modeling, as well as the aspects that contribute to the negative opinions of users through analysis of user ratings and sentiment. 
  
 \noindent\textbf{Results:} We identified sports apps that cover 16 sports types (e.g., Football, Cricket, Baseball) and 15 main functionalities (e.g., Betting, Betting Tips, Training, Tracking). We also extracted 14 topics from the user reviews, among which three are specific to sports apps (\textit{accuracy of prediction}, \textit{up-to-dateness}, and \textit{precision of tools}). Finally, we observed that users are mainly complaining about the advertisements and quality (e.g., bugs, content quality, streaming quality) of sports apps.
 %In the study, we understood the leading sports domain, market of interest and the algorithm establishment among applications. The review analysis helped us to distinguish three new categories specific to sports apps. We understood the most bothering factors for the sports users and how the developers can make use of the information to improve the user experience and application performance. Our study evaluates the equal importance of domain-specific study along with generalized study. We also point out how the studied factors about the domain, functionalities and analytical algorithms used over different topics of reviews obtained and the sentiment of reviews can be useful in improving the mobile app development cycles, thus improving the user satisfaction. \heng{be specific: explain how ``the results can be useful in ...'': either use summaries or examples}\bhagya{updated}

\noindent\textbf{Conclusion:}
%The results we listed can help the stakeholders to understand the sports market and the functionalities of the background study for their application. The main topics being discussed in the reviews and the sentiment associated with them give the real-time feedback, feature requests, and bug reports, along with sports specific metrics like the accuracy of prediction, up-to-dateness and precision of tools to improve the measures and hyper parameters being used in the applications.
It is concluded that analyzing a targeted category of apps (e.g., sports apps) can provide more specific insights than analyzing apps across different categories while still being relevant for a large number (e.g., tens of thousands) of apps. Besides, as a rapid-growing and competitive market, sports apps provide rich opportunities for future research, for example, to study the integration of data science or machine learning techniques in software applications or to study the factors that influence the competitiveness of the apps. %in the competitive market. 

\end{abstract}

%% - \Keywords{} with LaTeX markup, at least one required
%% - \Plainkeywords{} without LaTeX markup (if necessary)
%% - Should be comma-separated and in sentence case.
%\Keywords

\maketitle

% \input{texfiles/abstract}

% \begin{IEEEkeywords}
% Sports apps, user review, topic modelling, sentimental analysis
% \end{IEEEkeywords}

%\keywords{sports apps, user review, topic modelling, sentimental analysis}

\section{Introduction} \label{sec:intro}

\begin{comment}
\heng{Reorganize the Intro section based the comments below.}

\heng{First paragraph: 1) Importance of mobile app analysis (app market size is big; app store data enable app analysis); 2) Prior work typically either analyze general popular apps (give examples) from the entire store or focus on a small targeted set of apps (give examples), the former is too general (different categories of apps have very different functionalities) while the later is too specific. }

\heng{Second paragraph: 1) In this work we perform a analysis focus on an app category. We aim to find out whether such an analysis can provide more specific insights while still benefit a large number of apps in the category. 2) We choose the sports app category which is reportedly including over XX apps. 3) Give reasons why we choose the sports category.}

\heng{Fourth paragraph: 1) Summarize the goals, the data collection and analysis approaches. 2) Introduce the RQs}
\bhagya{addressed above comments}
\end{comment}

Mobile app stores (e.g., Google Play Store) provide a rich collection of apps for users to download and a wealth of information concerning users' feedback on the apps (e.g., user reviews).
Prior work has performed extensive studies on mobile app information that is available in app stores~\cite{Martin2017Survey}.
These studies often analyze a large number of apps across different app categories (e.g., \cite{chen2021should},\cite{hassan2022importance},\cite{gao2018online}
or focus on a small set of particular apps (e.g., \cite{gao2018online},\cite{frie2017insights} \cite{brown2020review}).
For example, Chen et al.~\cite{chen2021should} collected the reviews of 31,518 apps across all app categories of the Google Play Store (including the top 500 popular apps from each category), identified UI-related reviews, then manually analyzed a random sample of UI-related reviews to characterize UI-related issues.  
The findings of these types of studies are usually generalizable to different app categories. However, such findings may miss the particularity of a specific app category.
On the other hand, studies that focus on a small set of apps may derive findings that are too specific for the studied apps and lack generalizability. 

An app category in the app store usually contains thousands to hundreds of thousands of apps. For example, there are 46,625 apps in the \textit{sports} category in the Google Play Store~\cite{AppbrainStats}.
We assume that analyzing apps from such a targeted category can provide more \textit{specific} insights (i.e., specific to the app category) for \textit{a large number} of apps (e.g., tens of thousands of apps) in the app store. 
Thus, in this work, we aim to validate our assumption and perform a study of a large number of Android apps from a single category. %\textit{sports} apps in the Google Play Store. 
Specifically, we focus on the apps in the \textit{sports} category in the Google Play Store.

We focus our study on the \textit{sports} app category for several reasons. First, it is a large and fast-growing app category. It is reported that the \textit{sports} category includes over 46K apps~\cite{AppbrainStats}. In recent years, the \textit{sports} category has been growing very fast, with 35\% of the top 500 \textit{sports} apps released in the last three years.
It is predicted that the sports technology industry is going to have an average annual growth of 16.8\% and reach a market value of \$36.2 billion USD by 2028, and smartphone apps make up a key component in this industry~\citep{SportsTechMarketAnalysis}. Thus, we expect that \textit{sports} apps will continue to grow rapidly for at least a few years to come.
Second, although apps in the \textit{sports} category share some similarities, they cover a variety of sports domains (e.g., soccer or basketball) and functionalities (e.g., training or news), and widely adopt emerging technologies such as data analytics, artificial intelligence, internet of things, and social media integration~\citep{SportsTechMarketAnalysis}. Such variety enables our study to identify common issues across the entire app category, the ones that are specific to a sub-category of apps.

In this work, we study a carefully curated set of 2,058 \textit{sports} apps in the Google Play Store. As far as we know, no prior work has studied such a large number of mobile apps in a single app category. We exclude \textit{sports game} apps as they have very different characteristics from our considered \textit{sports} apps. 
Based on the top 500 free sports apps provided by AppBrain~\cite{AppbrainTopSports}, we extracted keywords related to \textit{sports} apps, then searched through the Google Play Store and manually verified each of the resulting apps. 
Our work combines qualitative analysis and quantitative analysis to understand the characteristics of the \textit{sports} apps, the topics in users' reviews, and the factors that users complain about these apps. Our work highlights the benefits of analyzing apps in a specific category and provides insights for \textit{sports} app developers to improve their apps or build new apps. Specifically, our work aims to answer the following three research questions (RQs).
%In this paper, we will closely analyze the characteristics of the market of interest, the context being discussed by the user via reviews and the factors the users are not happy about the sports apps. The data source for our research is Google Play store apps data and the reviews for all sports apps. For extracting sports apps, we leverage expert knowledge followed by review analysis using natural language techniques like topic modeling and sentimental analysis to understand the polarity of the data.We conclude our results from the mentioned analysis to aid both the sports app industry users, developers, and stakeholders. To infer the trends and patterns that can be useful for sports app development and maintenance, we have defined three research questions as follows.

%\heng{Below, for each RQ, give a one-sentence motivation, main approach, and key results.} \bhagya{updated}

\noindent $\bullet$ \textbf{RQ1: What are the characteristics of \textit{sports} apps?}
To understand the characteristics of sports apps, we manually study a statistically representative sample of sports apps. We observe that sports apps cover a wide spectrum of sports types (e.g., football) and functionalities (e.g., training). On the other hand, the sports app market is highly competitive, with many apps covering the same sports types and functionalities. 
%Our primary data source is the Google Play store; although the Google Play store has a category called sports, all sports apps are not categorized or labelled under it.This evaluation helps us to understand the distribution of data we have also it helped us to understand the sports market of interest, main app focuses and the algorithms being used by the current market. So to understand the categories of sports apps and classify them, we are manually analysing the sample set of data.
%Followed by creating a sample set of the data set and we categorize them based o three features, i.e., (1) Type of sports, (2) Main functionality, (3) Analytical/statistical algorithm used. Our results highlighted, Football as the most demanded market , betting tips and training as main functionality and the increasing using analytical algorithms in sports application.

%\Foutse{this sounds like the motivation for the question is just to understand the distribution of data...}. \heng{use concrete results instead of the motivational text (you know this without the study)}\bhagya{updated}

\noindent $\bullet$ \textbf{RQ2:  What are the topics raised by \textit{sports} app users in their reviews?}
%\heng{Motivation should include we want to examine if analyzing a specific app category can bring new perspectives.}\bhagya{added one line update}
%\heng{fix the missing citations later}
%App reviews are a direct path of communication between the users and the developers. \citep{hassan2018studying}. 
Understanding users' perceptions of the apps communicated in their reviews can aid in validating the quality of the apps~\cite{malavolta2015end,hassan2018studying}. 
In this research question, we apply topic modeling to extract topics from the user reviews of sports apps. % techniques~\citep{zevcevic2021user,hassan2022importance,Martin2017Survey,kalaichelavan2020people} to conclude the context of the discussion thus derive intuitions for enhancing the application software quality. 
%Our study examines the importance of specific category based analysis, and the nuances it can bring to the perspective of the data. We found 3 sports apps specific categories with respect to the new machine learning and sensor based techniques being used in sports apps.
%\heng{summarize key results}\bhagya{updated}
We derived 14 topics from the users' reviews of the 2,058 studied sports apps, including topics identified in previous review analysis studies, as well as three new topics that are specific for sports apps including \textit{accuracy of prediction}, \textit{up-to-dateness}, and \textit{precision of tools}.

\noindent $\bullet$ \textbf{RQ3: What do users complain most about \textit{sports} apps?}
%\heng{motivation missing} \bhagya{updated}
User ratings and reviews are indicators of how users feel about the apps. In this research question, we analyze the user ratings and the sentiment of the user reviews to understand the factors that users complain about sports apps. 
We observe that users are mainly complaining about the advertisements and quality (bugs, content quality, streaming quality) of the apps.
Besides, \textit{Streaming}, \textit{Betting}, \textit{Team management} and \textit{League management} are the most complained about sports app functionalities.

%In the previous research question, we categorized the different topics discussed in the reviews in general. In this research question, we focus on low sentiment reviews. Also, we use sentiment-rating mismatches to eliminate the inconsistent reviews that would create noise in the dataset.By focusing on understanding the complaining factors of the users and understanding the functionalities associated with them is to locate the issues associated with application software development to aid the developers. We compare our results to the other categories of sports and the factors the users are not happy about across different types of sports apps. Our results shows that most of the users are happy about the sports apps however the the complaining factors associated with topics are negative feedback and bug reports, those categories can be focused more by the developers to improve the user experience. \heng{summarize key results}\bhagya{updated}

Through our study of the 2,058 apps in the sports app category, we conclude that analyzing a targeted category of apps (e.g., sports apps) can provide more specific insights than analyzing apps across different categories while still being relevant for a large number (e.g., tens of thousands) of apps. For example, we identified more specific characteristics of the apps (e.g., sports types, functionalities) and new topics raised by users in their reviews that are specific to sports apps. \review{2-2}{In addition, we observe that sports apps are a rapid-growing and competitive market, which provides rich opportunities for future research, for example, to study the integration of data science or machine learning techniques in software applications (e.g., by exaiming the source code of these sports apps where data science or machine learning techniques are used), or to study the factors that influence the competitiveness of the apps in the competitive market (e.g., leveraging regression models to explain how the app features (e.g., readabilty of descriptions)  impact app ranks or ratings).}

To help future work replicate or extend our work, we share our replication package\footnote{Replication package: https://github.com/mooselab/Sports-Apps-Analysis, Password for Dataset: Sports123!}.

%\heng{Organization of the rest of the paper.}
\noindent \textbf{Paper organization.} The remainder of the paper is organized as follows. Section \ref{sec:setup} describes the experimental setup of our study. %overall data collection and analysis. 
Section \ref{sec:results} describes our results from the experiments conducted for each research question.  Section \ref{sec:discussions} discusses the implications of our findings. The threats to the validity of our findings are discussed in section \ref{sec:threats} and followed by the related works in section \ref{sec:related}. Finally, we conclude our paper in section \ref{sec:conclusions}.

\section{Experiment Setup} \label{sec:setup}

%\subsubsection{Overview}
%\heng{Use an overview figure with boxes and arrows to illustrate the overall study. The figure should make it easy to correspond the other subsections/subsubsections to the steps in the figure. Include numbers in teh stesp (e.g., number of apps before/after filtering, number of collected reviews, etc.). Check Fig. 1 in this paper: https://www.hengli.org/pdf/Hassan2022PeerApps.pdf}

\begin{figure*}[t]
    \centering
    \includegraphics[width=.85\textwidth]{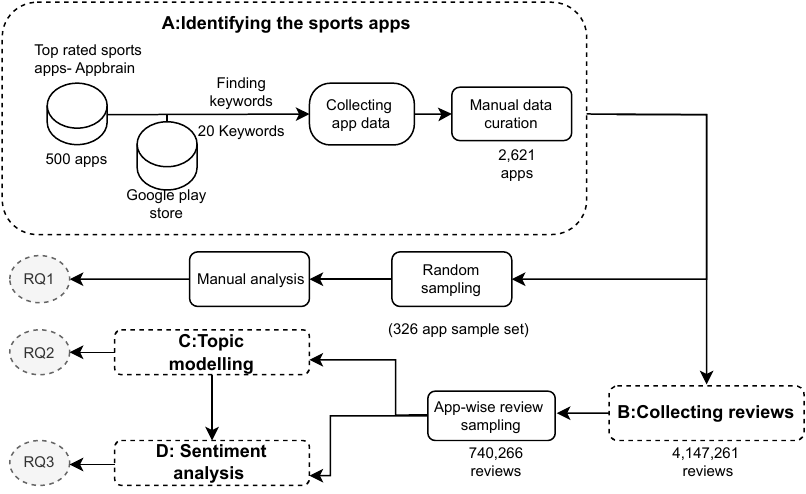}
    \caption{Overview of our study.}
    \label{fig:overview}
\end{figure*}

%This section explains the processes that we followed to create, evaluate, curate, and analyze data. 
Figure~\ref{fig:overview} shows an overview of our study.
This section explains our process of identifying, collecting, and preparing the data for answering our research questions.
%The detailed steps are described in the rest of this section.
The detailed approaches used for answering each RQ are described in Section~\ref{sec:results}. 

\subsection{Identifying the Sports Apps}
%\heng{Include the steps of identifying apps in the overview figure.}

%\heng{Reorganization as commented below:}

%\heng{Include these steps (for identifying sports apps) in this subsection: Identify sports apps from AppBrain -> Manually filtering of sports apps (removal of sports game apps) -> Identifying keywords for sports apps -> Searching apps from Google Play Store -> Manual filtering of sports apps}

Google Play store provided category tags for each app. However, from our preliminary analysis, we understood that developers placing their apps in the play store do not pay much attention to the tags available. \review{2-4.1}{
The category tags available in the play store are not reliable as a solo criterion for selecting a specific category of mobile apps from the Google Play Store \cite{al2016clustering}. Google Play Store developers are required to select a genre for their apps upon uploading them. Consider and examples of the apps, RealFevr (com.realfevr.fantasy) is a fantasy sports game app which is classified as sports app instead of sports game app.Similarly, PlayersOnly - Sports Social (com.playersonlymobileapp), which is a social media app for athletes that contains sports-specific information categorized into social instead of sports \cite{googleplay}. The sports app industry is significantly broader and encompasses areas such as fitness apps, sports tools, and dedicated social media platforms for sports, among others, as indicated by a citation from Statista \cite{statista}. Consequently, relying solely on the sports category provided by developers on the Google Play Store may not accurately encompass the comprehensive range of sports applications available.}

% Thus the category tags available in the play store are not reliable for selecting as a solo metric for data stratification of mobile apps from Google Play store \cite{al2016clustering}. Google Play Store developers are required to select a genre for their apps upon uploading them. While Google's recommended sports category encompasses aspects such as "Sports news and commentary, score tracking, team management, game coverage" \cite{googleplay}, the sports app industry is significantly broader and encompasses areas such as fitness apps, sports tools, and dedicated social media platforms for sports, among others, as indicated by a citation from Statista. Consequently, relying solely on the sports category provided by developers on the Google Play Store may not encompass the comprehensive range of sports applications available \cite{statista}}.  Besides, the Google Play store only returns a limited number of apps for each search query (e.g., searching the keyword ``sports'' return fewer than 250 apps). To identify sports apps from all the other apps, we need to have a set of keywords that we can use as the search keywords. 
%\newline 
To address these challenges and identify a large number of \textit{sports} apps, we started with an initial set of \textit{sports} apps, based on which we identified a set of related keywords. Then, we searched these keywords in the Google Play store and finally manually verified each of the resulting apps.
The approach we used to generate those keywords is to collate the description associated with the apps and thus look into the contextually important words to use as the search words for the next iteration. 
%\heng{Rephrase: briefly summarize the steps before digging into the details of each of them}
%\bhagya{modified}

\subsubsection{Top-rated sports apps from AppBrain}
AppBrain\footnote{https://www.appbrain.com/} is one of the leading companies that collect information regarding the Android app ecosystem. From AppBrain, we downloaded a dataset of top-rated 500 free apps\cite{AppbrainTopSports} in the US region\footnote{https://www.appbrain.com/apps/popular/sports/}. \review{2-4.2}{According to the analysis of mobile app usage in 2020, the US had the highest number of sports app downloads \cite{stateofmobile}. Thus, selecting the US market allows us to collect a large amount of user feedback data (reviews and ratings)}. From each of these apps, we extracted the following metadata: Google Play URL, Short description, Launch time, Last update time, Rating, and Comment count. % as the meta data that we will analyse to understand the market and trends of the data. 
We used this set of \textit{sports} apps as our reference set, and we used the Google Play %\heng{Google Play or play: make it consistent throughout the paper}\bhagya{Google Play -addressed} 
URL to get a detailed description of these apps for our next step. 

%\heng{Include collecting app data steps in this subsection: collecting app ratings (used to RQ3), reviews (used for RQ2 and RQ3), descriptions (used for RQ1). }

%The first step is to get the sports apps from the Google Playstore and for that we need appropriate keywords to search for.
\subsubsection{Finding keywords for sports apps %iterative search\heng{Iterative identification of keywords?}
}
%\bhagya{keyword search was one iteration, but used the keywords to find more apps by iterative search in the playstore}
As using the keyword ``sports'' may retrieve many false positives (e.g., sports games) and the Google Play store limits the number of apps retrieved per request to 250, we need to identify a set of keywords which are used to search for \textit{sports} apps in the store. %So we had to find a list of keywords to run iterative queries, extract a maximum number of apps, and remove the duplicates. 
\review{3-3}{We looked into the title and description of the top-rated sports apps and applied the keywords extraction algorithm YAKE \citep{campos2018yake} with a window size of 3. \review{2-4.3}{The window size was set to 3 based on established practices in extracting the most significant keywords \cite{harman2012app}. Additionally, the analysis employing multiple local features with YAKE indicated that a window size of 3 provides a balance between complexity and the inclusion of valuable information \cite{campos2020yake}.} YAKE is an unsupervised keyword extraction method which provides a score for the contextually important keywords in a corpus. We select the most important keywords, which can contain a maximum of 3 keywords.
Each of the extracted keywords underwent a manual analysis to affirm its relevance to \textit{sports} apps.}
%and used these keywords for further iterative search and extracted apps related to the keywords.
We performed an iterative process to identify the keywords: the titles and descriptions of the apps searched from the identified keywords are further used to extract new keywords. 
We stopped the iterative process until no new sports-related keywords could be identified.
%\heng{Rephrased the above process, check if it is right.} \bhagya{addressed}

The step resulted in a final list of 20 %\heng{or 21?}\bhagya{20, one of them was substring so removed it} 
sports-related keywords, including \textit{golf GPS, sports betting, sports betting tips, soccer betting,
fitivity, fitness,
football,
betting tips,
soccer training,
sports leagues,
football tips,
NBA League Pass,
live sports,
sports network Live,
Pro sports subscription,
state football,
football training,
champions league,
premier league,
sports player.}

\subsubsection{Collecting app data} \label{sec:collect-app-data}
\review{3-4}{Using the keywords identified from the last step, we obtained 2,621 apps by searching in the Google Play store after removing duplication. Then, we extracted the Metadata of each app using Selenium HTTP requests. We extracted the following Metadata: The App name, Description, Rating score and Downloads are used in our RQ1 to study the characteristics of the sports apps. The Google Play URL is used to extract the app reviews that are used to answer our RQ2 and RQ3 and Ads status for comparing our findings to market scenarios.}
%We did the iterative extraction of apps with these keywords, and our final dataset was 2621 apps. The iterative extraction was done using Selenium. \heng{should the iterative process described in the keyword identification step?}
%We used the keywords as the search keyword in the URL and connected them to the Selenium HTTP request. 
%--2--Using the keywords identified from the last step, we obtained 2,621 apps by searching in the Google Play store after removing duplication. 
%Once we got the entire URL of the Apps, 
%--2--Then, we extracted the Metadata of each app using Selenium HTTP requests. We focused on the following Metadata: Google Play URL, App name, Providers, Tags, Ads status, Description, Rating score, Rating Count, Downloads, Size, Date of update, Content rating permission, and Cost. %However in this research we are focusing on few of the metadata such as content rating permission to understand the user groups, ads status to evaluate the concerns from the users for ads in the apps. We kept the rest of the details in our data set for further exploration in the future.
%--2--The App name, Description, Rating score and Downloads are used in our RQ1 to study the characteristics of the sports apps. The Google Play URL is used to extract the app reviews that are used to answer our RQ2 and RQ3. 
% \heng{check if this is correct; remove the ones from the list that are not used in this paper.}

\subsubsection{Manual data curation}

Although we collected the 2,621 apps using sports-related keywords, some of the resulting apps may not be \textit{sports} apps. For example, some \textit{sports game} apps were included in the search results. %the chances of non sports apps or games apps to be in our data is not 0. 
Thus, we performed a manual verification of each of the resulting apps to remove false positives.   %on confirming the apps that we have is completely sports apps. 
In this manual process, we excluded all game apps and apps that are not related to sports, such as general streaming apps (e.g., Spotify). Our final verified dataset contains 2,058 \textit{sports} apps.

\subsection{Manual analysis}\label{sec:sample} %\heng{This is not part of identifying sports apps? It should be a main step?} \bhagya{modified}
From our final set of 2,058 apps, we randomly sampled a set of 326 apps, which represent a confidence level of 95\% and a margin of error 5\%. The apps in the random set are manually examined to understand the characteristics of \textit{sports} apps (i.e., RQ1), including their sports domains, main functionalities, and their used analytical methods. The details of our manual analysis process are described in Section~\ref{sec:rq1}.
%We created a sample set of the curated data to understand them in more discrete level. We did manual analysis on the sample dataset to understand 3 features i,e (1)Type of sports (2) Main functionality (3) Analytical/statistical algorithm used. 

\begin{table}[!t]
\centering
\caption{Five-number summary statistics on a number of reviews. %\heng{the second column is number of reviews? Also add the mean value. Rotate the table to have two rows only to save the space}\bhagya{I thought 50th percentile is mean value!}\heng{50 percentile is median, not mean; always separate every three digits when reporting numbers}\bhagya{noted}
}
\label{table-review-count-total}
\begin{tabular}{|l|l|l|l|l|l|}
\hline
\textbf{Percentile Distribution} & 0 & 25 & 50 & 75 & 100 \\ \hline
\textbf{Number of sampled reviews} & 0 & 7 & 41 & 294 & 3398 \\ \hline
\textbf{Number of collected reviews} & 0 & 7 & 42 & 322 & 204,970 \\ \hline
\end{tabular}
\end{table}

\subsection{Collecting reviews}
We used the google-play-scraper API\footnote{https://pypi.org/project/google-play-scraper/} %\heng{citation}\bhagya{couldnot find citation added as footnote} 
to extract the reviews of our final set of 2,058 \textit{sports} apps. %, and we extracted reviews associated with all ratings with all sports we are analyzing. 
The API returned the following data we considered in this study: reviews and rating scores associated with them. \review{2-6}{A total of 4,147,261 reviews were extracted encompassing all the apps under consideration. However, in order to address potential imbalances within the dataset (e.g., an app can account for up to 5\% of all the reviews) and comprehensively represent the population of apps, we conducted an app-wise sampling approach with a confidence level of 95\% and an error margin of 5\% (i.e., we calculated the sample size for each app individually). This sampling method effectively accounts for variations in the number of reviews across different apps while maintaining the integrity of the population characteristics. Consequently, we sampled a subset of 740,226 reviews, encompassing all the apps, as a representative sample for further examination and inference. Table 1 provides a five-number summary of the number of reviews of each app before and after the sampling process.
}
Table ~\ref{table-review-count-total} shows five-number summary statistics and the average number of reviews of all apps that we study.%\heng{add a table showing the five-number summary statistics and the average number of reviews}\bhagya{will this be a repeated info from the distribution reviews over different topics Tab 3?}\heng{the distribution here is about the number of reviews distributed over apps, Table 3 is about (percentage) distribution over topics.} \bhagya{updated}

\subsection{Topic Modeling}\label{tp:tm}
To understand the topics raised by sports app users in their reviews, we perform topic modeling to extract semantic topics from all the collected reviews. Then, we analyze the resulting topics to answer our RQ2. The details of our topic modeling and analysis approaches are described in Section~\ref{sec:rq2}.
%The automated topic modeling algorithms used to understand the main focuses of the reviews that are being studied.Followed by manual analysis on topic selection and topic naming
%\bhagya{I added this section here as the diagram includes this. But more details on topic modeling are already present in the RQ so basic introduction of the idea is okay here??}

\subsection{Sentiment analysis of user reviews} 
We perform sentiment analysis of all the reviews of the collected \textit{sports} apps to understand user-perceived negative opinions regarding the \textit{sports} apps (i.e., RQ3). In particular, we analyze users' sentiment in the reviews associated with each topic derived from our topic modeling process. The details of our sentiment analysis approach are described in Section~\ref{sec:rq3}.

%\heng{Move the following content to RQ3 approach and rewrite it (currently not in a good form).} \bhagya{whole 2.5?}\heng{as sentiment analysis is specific to RQ3, use a similar format as for topic modeling and RQ2 approach: briefly mention the sentiment analysis step in experiment setup, then describe the details of the approach in RQ3-approach (in an organized way)} 

%\heng{this comment not fixed yet: please move the details of RQ3 approach to RQ3, similar as topic modeling for RQ2.}

%\newline

% \balance{
% \bibliographystyle{ACM-Reference-Format}
% \bibliograph{sports-apps}
% }

%\input{texfiles/methodology}

\section{Experiment Results} \label{sec:results}

\subsection{RQ1: What are the characteristics of sports apps?} \label{sec:rq1}

\subsubsection{Motivation} 
It is reported that there are over 46K apps in the \textit{sports} category in the Google Play Store~\citep{AppbrainStats}.
However, the characteristics of these \textit{sports} apps (e.g., their provided functionalities) are not clear.
%Executing app analysis with a focus in the sports category is an unique study which has never performed on. 
Therefore, we first investigate the characteristics of these apps, which can provide us with better context when analyzing these apps in terms of their review topics in RQ2 (Section~\ref{sec:rq2}) and user-complained aspects RQ3 (Section~\ref{sec:rq3}). In particular, in this RQ, we investigate the types of these apps, their main functionalities and the analytical algorithms used in them. Our results can help researchers and practitioners better understand the domain of \textit{sports} apps and provide context for our further analysis in RQ2 and RQ3.
%When an application being launched in Google Play store there are a set of data being associated. We incorporated those details to the dataset that we generated and did a preliminary exploratory study on the data. This will help us to understand the market of interest in the sports mobile application industry, the growth trends, and the user interactions by analysing the distribution of user ratings and downloads. We concludes our study by evaluating whether the apps in the app store reflects the real-time interest towards the type of sports and the requirements of the sportsmen and/or sportsfans with the development patterns and algorithms being used as a reference to the developers and the researchers.

\subsubsection{Approach} 
In this RQ, we first study the store characteristics of \textit{sports} apps, including their over-trend and popularity (in terms of downloads and ratings). Then we manually examine their intrinsic characteristics (non-store characteristics), including their targeted sport types, their main functionalities, and their used analytical algorithm types.

%Once we extracted the Metadata for our selected 2058 apps, we started our analysis to answer our RQ1. However, the data followed by cleaning the exact representation for downloads, ratings, etc. The manual labelling of data follows this to understand three characteristics: Type of sports, main functionality and algorithms used.
%In the  exploratory data analysis the we were trying to answer the following metrics a) The trend of the apps being launched in the playstore b) The distribution of the ratings, review count c) The categorisation of the sports apps in three facets : the type of sports, main functionality of the apps and the analytical algorithms types being used in the apps.

% \heng{In the first paragraph, briefly summarize what analyzes are performed. Then discuss the details in the following.}
% \bhagya{I have combined the 3 sections into one as it is only one line for each section except the methods}

\noindent\textbf{Overall trend of \textit{sports} apps.} 
%\heng{Describe why this is important, what data you use, how it is analyzed.}
%\bhagya{need feedback on the why importance part}
We study the overall trend of sports apps to understand the growth of the sports app industry.
%This analysis to understand how extend the growth in the sports app industry has been in last decade. 
%This data would help us to validate why it is important to choose the category and the importance of our research to the future. 
We looked into the launch time of the 500 top-rated apps from Appbrain and counted the number of these apps created each year. We did not consider all the 2,058 apps because the Google Play Store does not provide the launch time of the apps. Thus, we limited our analysis to the top 500 apps from Appbrain, which provides the launch time. The launch time of these 500 apps ranges from 2010 to 2020 (the time when the data was collected).

\noindent\textbf{Downloads and ratings of \textit{sports} apps.} 
We study the number of downloads and the ratings of the sports apps to understand their popularity and user-perceived quality.
%As part of the exploratory study of the metadata that we extracted the distribution of downloads and ratings gave us the understanding of how much user engagement we have for apps. 
%Although, the downloads and ratings are highly correlated the further studies on rating in RQ3 helps us to differentiate the more information it contains. 
We considered all the 2,058 manually verified sports apps and extracted their download and rating information from the Google Play Store (see \ref{sec:collect-app-data}). In the Google Play Store, the average rating of an app is provided as a real number ranging from 1 to 5 (e.g., 4.5). The number of downloads is provided as a range value (e.g., 100K+ or 1M+). We analyze the distributions of the number of downloads and ratings of the apps in this RQ.
% considered buckets such as less than 1000, 10k , 100k , 1M+
%\heng{Describe why they are important, what data you use, how it is analyzed.}
%\bhagya{updated}

%\noindent\textbf{Categorization of the apps by sports type, main functionality and algorithms used.}
\noindent\textbf{Categorization of \textit{sports} apps.}
To understand the characteristics of sports apps and provide context for our further analysis, we manually examined a statistically representative sample from the 2,058 apps to derive a categorization of these apps from three perspectives:
\begin{itemize}
    \item \textbf{Type of sport}. The target sport of the app, such as \textit{football} or \textit{golf}. An app is labelled as \textit{general} if it does not target a specific type of sport. In case an app does not associate with any kind of sports but a generic purpose e.g., news app we classify it to \textit{NA}.
    \item \textbf{Functionality}. The main functionality of the apps, such as \textit{training} or \textit{betting}. However, in An app with multiple functionalities, we only consider the primary functionality. \review{2-4.4}{The main functionality of the apps, such as training or betting. However, in An app with multiple functionalities, we only consider the primary functionality. Our objective is to examine the relationship between these topics among applications that offer similar functionality, aiming to identify prominent topics among competing applications. Consequently, it becomes crucial to prioritize the main features of these applications in order to identify distinct attributes that are specific to each functionality. By avoiding the classification of multiple functionalities together, we can minimize noise and ensure a focused analysis at the functional level, particularly in the context of competition among similar applications. For example, app Yahoo sports describes the app as \textit{``Get sports news, scores, live results \& updates on Yahoo Sports so you don’t miss a second of the action. Dig into analysis, prediction and commentary from our expert editorial writers about football, basketball, baseball, and more.''} The primary functionality is Live updates, and secondary could be betting tips as they provide analytics. Whereas consider app Football betting tips it says \textit{``SuperTips' App Features: Up-to-date football betting tips every day, Football tips from the most successful tipsters, Match statistics and live scores''}where bettting tips is the main functionality and live scores are secondary. }
    \item \textbf{Type of analytics}. We observe that many sports apps use statistical or predictive analyses to achieve their goals (e.g., predicting match outcomes). Thus we categorize the type of analytics used in the apps. We use an ``unknown'' label when we could derive the type of analytics used in an app. Similarly we used ``NA'' if the application does not use any kind of analytical algorithms used in them. \review{2-7}{App descriptions did contain specific keywords such as "predictive" or "statistical models" that indicated the presence of such analyses. For example, the Footbe app (net.footbe.footbe) describe the app as “The soccer results prediction app that makes sense out of stats and performances for you. Footbe provides in-depth soccer analytics, predictions, and betting tips driven by machine learning. We developed unique predictive algorithms based on the player rating system.” Simlarly, the Ultra tip bet app (com.ultrabt.fdykcs) describes “Ultra Tips Bet is built on a unique prediction algorithm that's based totally on player rating systems, making it stand out from the rest. Ultra Tips Bet's prediction algorithm is based on data from past matches, group and player performance, and other factors such as climate and injuries. This data is analyzed and processed using machine getting to know algorithms to generate a unique prediction for each game.”} 
\end{itemize}
We created a sample set of 336 apps with a confidence level of 95\% and a confidence interval of 5\%. We collated the Application name and the description associated with it as the information needed to be examined. If the data is insufficient, we direct to the play store links to verify more details.
The manual labelling was done in 3 stages by researchers with a software engineering background.
\begin{itemize}
    \item First round, three researchers labelled 10\% of  apps together until we reached a common understanding.
    \item Secondly, the first author labelled 70\% and then discussed with other researchers and reached a consensus.
    \item Finally, two researchers labelled 30\% independently to measure the reliability. A consensus was reached for each label.
\end{itemize}

 We used Cohen's kappa score~\cite{cohen1960coefficient} to measure the inter-rater reliability. For the categorization of the sports type and the functionality, we achieved Cohen's kappa score of 1, which indicates \textit{perfect} reliability~\cite{mchugh2012interrater}.
 For the type of analytics, we achieved a Cohen's kappa score of 0.81, which indicate a \texttt{strong} reliability~\cite{mchugh2012interrater}. 
 %The final agreement score was high\heng{give a specific number}, which indicate a \heng{high/acceptable/etc/, depending on the interpretation of Cohen's Kappa} reliability\heng{cite paper for the interpretation}.
 %The conclusion was made by discussing every conflict and finalising the agreement
 
 %\heng{do not mention the specific types here as you have not explained them yet (they are in the results); just mention the agreement score is fine} 
%\heng{Explain that the agreement scores indicate a high reliability, cite a reference.}
%\bhagya{updated}

%Here we are trying to understand the three questions first what are the users most interested sports type, and secondly what are the app market in place related to sports and finally to check the algorithms that has been used in the apps to understand the adoption of Machine learning related practices in the sports apps industry.
%\heng{Clearly describe the steps of the categorization, what data is used (and what information you are looking for), how many people are involved, what agreement ratio is used and how. Example can be found in this pape (Section 3.1.4): https://www.hengli.org/pdf/Li2020LoggingQualitativeStudy.pdf}
 %\bhagya{The labelling is quite straight forward for except for algorithms used so multilple iterations was not necessary as in the paper mentioned so confined them in one paragraph}

\subsubsection{Results} 
% \input{texfiles/images/category-label-table}
%\heng{Organize the results section this way: Each paragraph is about one take home message. At the beginning of each paragraph, use a bold sentence to summarize the take home massage. Then use the data (with illustration of figure/table) to support/explain the message. Finally, discuss the implication of the finding (go beyond the results and discuss what's the implication for practitioners/researchers/rest of the paper).}

%\heng{Discuss the categorization results before other parts, so you can discuss other characteristics (trends, ratings, etc.) considering the different app types.}
%\bhagya{This cannot be done as in the trend and categorisation is done on different datasets - data from Google Playstore does not have a launch date}

% \begin{figure}[]
% \begin{subfigure}[b]
% \centerline{\includegraphics[width=0.25\textwidth]{texfiles/images/trend.png}}
% \caption{Trend of number of apps launched per year}
% \label{fig:app-trend}
% \end{subfigure}
% \begin{subfigure}[b]
% \centerline{\includegraphics[width=0.25\textwidth]{texfiles/images/rating-score.png}}
% \caption{Distribution of rating score}
% \label{fig:app-trend}
% \end{subfigure}

% \end{figure}

%\begin{figure}
    %  \centering
     %\begin{subfigure}[b]{0.49\textwidth}
    \begin{figure}[!t]%{0.49\textwidth}
         \centering
         \includegraphics[width=.85\textwidth]{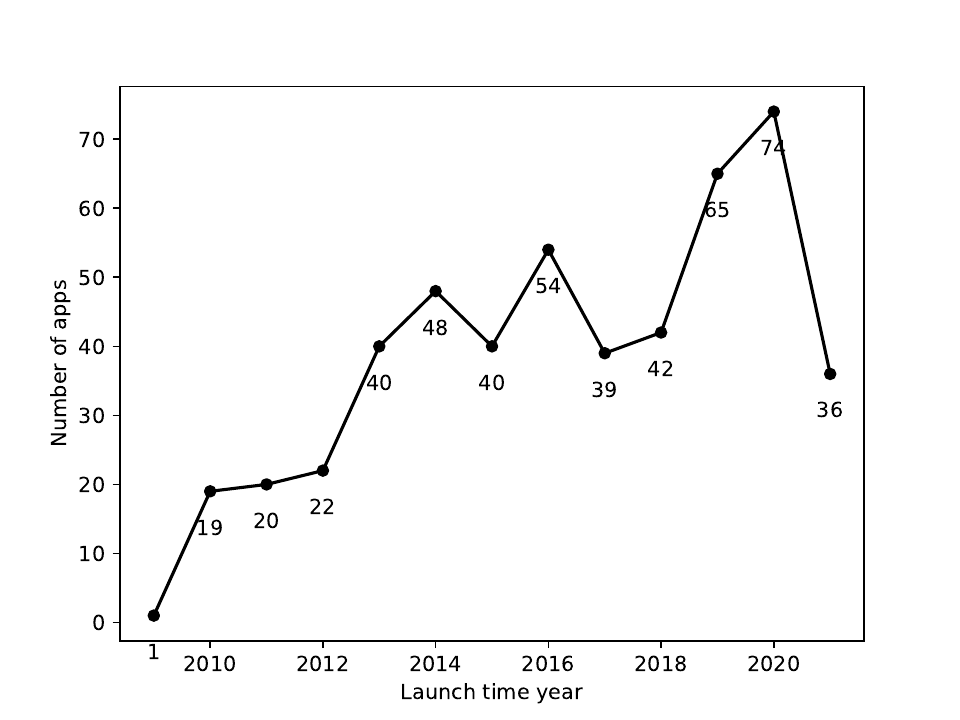}
         \caption{The number of sports apps launched per year in Appbrain's top-500 list.}
         \label{trend}
     %\end{subfigure}
    \end{figure}

     %\hfill
     %\begin{subfigure}[b]{0.49\textwidth}
     % \begin{figure}[!t]%{0.49\textwidth}
     %     \centering
     %     \includegraphics[width=.85\textwidth]{imgs/Rplot-rating-distribution.pdf}
     %     \caption{The distribution of rating scores of the studied sports apps.}
     %     \label{rating-distri}
     % \end{figure}
     % %\end{subfigure}
        %\caption{Metadata Analysis }
        %\label{twograph}
%\end{figure}

% \usepackage{booktabs}
\begin{table}[!t]
\centering
\caption{The distribution of the number of downloads of the studied sports apps. }
\label{table-downloads}
\begin{tabular}{|l|l|l|l|l|l|}
\hline
\textbf{\begin{tabular}[c]{@{}l@{}}Number of \\ downloads\end{tabular}} & \textless{}= 1000 & \textless{}= 10k & \textless{}= 100k & \textless{}= 1M & 1M+ \\ \hline
\textbf{Percentage of apps} & 13.08\% & 30.79\% & 8.85\% & 34.63\% & 12.65\% \\ \hline
\end{tabular}
\end{table}

\textbf{Sports apps have been increasing steadily in recent years.}
Figure~\ref{trend} shows the number of apps in Appbrain's top-500 list that were launched each year. 
The figure shows that the number of apps launched each year is continuously increasing from 2010 to 2020, with some minor fluctuations in the years 2015 and 2017.
%To understand the trends of the sports apps introduced in the Google Playstore, we looked into the top-rated 500 apps. We evaluated the apps trends of a decade. (Fig~\ref{trend})
The trend shows that the apps launched per year improved almost 4 times from 2010 to 2020. %in the sample size we evaluated.
Although the results only indicate the trend of the apps in Appbrain's top-500 list, we can infer that the overall number of sports apps may follow a similar trend.
Interestingly, the trend approximately matches the prediction that the sports technology industry is going to have an average annual growth of 16.8\% ($(1+16.8)^{10}=4.73$)~\cite{SportsTechMarketAnalysis}. It is reported that there are over 46K sports apps at the time of writing this paper~\citep{AppbrainStats}. 
\review{2-8}{ To have a more generic view, we did a comparison against the sports apps and non-sports apps according to the stats given by Data.ai\cite{appannie2021}. We found sports apps have grown significantly in recent years, marked by a 60\% surge in downloads and a 35\% upsurge in consumer spending between 2016 and 2020. However, non-sports apps have also experienced noteworthy growth during this period, with overall app downloads increasing by 45\% and consumer spending increasing by 20\%.}
%This improved trend in sports apps 
The fast-growing trend motivates us to focus our studies on sports apps. %Moreover, our further study focused on this category will help the developers and the stakeholders to understand the market and the user engagement better.
%\heng{use cross-reference as the example shown here}
%\heng{Discuss the implication: what's the implication for developers/researchers/rest of the paper? For example: This finding motivate us to focus our study on the sports app category.}
%\heng{Use a boxplot figure to show the distribution of the ratings: use one boxplot for each sport type; all boxplots combined in the same figure.}
%\bhagya{addressed}

%\textbf{Most users like the sports apps}
\textbf{Sports apps have a large number of user groups.} \textbf{Users are generally positive about the quality of the sports apps, with the average rating of the sports apps higher than general apps in the Google Play Store.} Table \ref{table-downloads} shows the distribution of the number of downloads of the sports apps. 47.28\% %(\heng{?}\bhagya{added values of <=1M and above 1M}) 
of the studied apps have more than 100K downloads, while only 13.08\% of the studied apps have fewer than 1000 downloads. %As we analysed previously the number of apps introduced recently are high thus the number of downloads can be also affected for the new apps. 
%However, 
%The apps with less than 1000 downloads are of only 13.08\% of our total extracted apps. 
%As there are a significant number of users that using sports apps, a specific study like this to understand the trends of users behaviour through reviews. 
%\bhagya{box plot is not added as this value seems to be added by the developers rather than the actual count - because in the dataset it was a mix of different values lie 1M , 1000000, 1M+ ,1000000+ etc}
Prior work emphasizes the importance of user ratings of mobile apps, reporting that 
69\% of users consider app rating is very important~\cite{hassan2018studying}.
Figure~\ref{rating-distri} shows the distribution of the average ratings of the studied sports apps.
Out of the 2,058 sports apps that we evaluated, 40\% of them are rated above 4.5, while only 0.42\% of them have ratings of less than 2. 
The average and median ratings of these apps are 4.0 and 4.1, respectively.
The result indicates that users are generally positive towards the majority of these sports apps. %the most of the apps in the market are very much of the interest to the users(Fig \ref{rating-distri}). 
%User ratings are the way the users conveys their attitude towards this category of apps. This value can be monitored to measure the user group interaction for developers and the stakeholders.
% \begin{figure}[h!]
% \centerline{\includegraphics[width=0.5\textwidth]{texfiles/images/rating-score.png}}
% \caption{Distribution of rating score}
% \label{fig:app-trend}
% \end{figure}
%When we did the median user rating for each sports type, the least user rating is 3.05 for base ball and the highest rating hockey with 4.77 and cricket and fitness are the categories with 4.5 ratings. Similarly the apps with functionality healthtip has the highest rating of 4.6, league management, betting tips and new are second with rating of 4.3. The functionality the users are not happy about is the tools. 
% \heng{Discuss the ratings across different sport type.}
% \heng{Compare the results with other papers showing ratings for general apps.}
Prior work on general sports apps reported that the average rating given for apps is 3.90 \cite{Martin2017Survey}. Thus sports apps have higher average ratings than general apps. 
%also previous studies confirms the importance of user ratings \cite{hassan2018studying} that 69\% users considers that app rating is very important.

%Finally we are addressing the important question to understand the top categories of sports, main functionality and the analytical and statistical algorithms used. In Table 1, we concluded the types of sports in the extracted apps.
Table~\ref{rq1results} shows our categorization of the sports apps in terms of the type of sport, the functionality, and the type of analytics. 
%The reliability of our categorization was done by cohen kappa score. For the Sports type and Main functionality, the Cohen Kappa score was 1. In contrast, in the algorithms used, the main mismatch was for three labels: Predictive analysis, Analytical statistics and Analytical statistics and Predictive analysis. We calculated the Cohen Kappa score for the mismatched items, which was 0.81. 

% \usepackage{booktabs}
% \usepackage{multirow}

\begin{table*}[!t]

\caption{Characteristics of the manual analyzed sports apps: type of sports, main functionality, analytics approach.} 
\label{rq1results}
\scalebox{0.415}{
\Large
\begin{tabular}{ccccccccc}
\textbf{\begin{tabular}[c]{@{}c@{}}Classification\\  category\end{tabular}} & \multicolumn{8}{c}{\textbf{Type of sports}} \\ \hline
Labels & \textbf{Football} & General & Fitness & Golf & Basketball & \begin{tabular}[c]{@{}c@{}}Car \\ Racing\end{tabular} & \begin{tabular}[c]{@{}c@{}}Martial \\ Arts\end{tabular} & \begin{tabular}[c]{@{}c@{}}American\\  Football\end{tabular} \\
Distribution & \textbf{36.16\%} & 23.73\% & 15.25\% & 9.04\% & 5.08\% & 1.69\% & 1.69\% & 1.13\% \\
Labels & Baseball & Cricket & Athletics & \begin{tabular}[c]{@{}c@{}}Football \\ and \\ Basketball\end{tabular} & Game & Hockey & Wrestling & NA \\
Distribution & 1.13\% & 1.13\% & 0.56\% & 0.56\% & 0.56\% & 0.56\% & 0.56\% & 1.13\% \\
\multicolumn{1}{l}{} & \multicolumn{1}{l}{} & \multicolumn{1}{l}{} & \multicolumn{1}{l}{} & \multicolumn{1}{l}{} & \multicolumn{1}{l}{} & \multicolumn{1}{l}{} & \multicolumn{1}{l}{} & \multicolumn{1}{l}{} \\
 &  &  &  &  &  &  &  &  \\
\textbf{\begin{tabular}[c]{@{}c@{}}Classification\\  category\end{tabular}} & \multicolumn{7}{c}{\textbf{Functionality}} & \textbf{} \\ \cline{1-8}
Labels & \textbf{Betting Tips} & \textbf{Training} & \begin{tabular}[c]{@{}c@{}}Live \\ updates\end{tabular} & Streaming & Tools & Tracking & Betting &  \\
Distribution & \textbf{24.86\%} & \textbf{24.86\%} & 15.25\% & 10.73\% & 7.34\% & 4.52\% & 3.39\% &  \\
Labels & News & \begin{tabular}[c]{@{}c@{}}Social \\ Network\end{tabular} & \begin{tabular}[c]{@{}c@{}}League \\ Management\end{tabular} & Radio & \begin{tabular}[c]{@{}c@{}}Team \\ management\end{tabular} & \begin{tabular}[c]{@{}c@{}}Health \\ Tips\end{tabular} &  &  \\
Distribution & 2.82\% & 2.26\% & 1.13\% & 1.13\% & 1.13\% & 0.56\% &  &  \\
\multicolumn{1}{l}{} & \multicolumn{1}{l}{} & \multicolumn{1}{l}{} & \multicolumn{1}{l}{} & \multicolumn{1}{l}{} & \multicolumn{1}{l}{} & \multicolumn{1}{l}{} & \multicolumn{1}{l}{} & \multicolumn{1}{l}{} \\
 &  &  &  &  &  &  &  &  \\
\textbf{\begin{tabular}[c]{@{}c@{}}Classification \\ category\end{tabular}} & \multicolumn{6}{c}{\textbf{Analytical /statistical algorithm used}} & \textbf{} & \textbf{} \\ \cline{1-7}
Labels & \begin{tabular}[c]{@{}c@{}}Predictive \\ analysis\end{tabular} & \begin{tabular}[c]{@{}c@{}}Analytical \\ statistics\end{tabular} & \begin{tabular}[c]{@{}c@{}}Descriptive\\  statistics\end{tabular} & \begin{tabular}[c]{@{}c@{}}Analytical statistics and \\ Predictive analysis\end{tabular} & Unknown & \textbf{NA} &  &  \\
Distribution & 18.08\% & 7.34\% & 7.34\% & 3.95\% & 1.13\% & \textbf{62.15\%} &  & 
\end{tabular}}
\end{table*}

%\textbf{Most famous sports apps category is Football}
\textbf{Sports apps cover a wide spectrum of sports, with football\footnote{Football refers to association football or soccer in this paper.} being the most frequently targeted sport type. However, the distribution of sports apps is disproportional to the popularity of sports.}
From Table \ref{rq1results}, we can understand that football (36.16\%), fitness (15.25\%), golf (9.04\%), and basketball (5.08\%) are the most common types of sports targeted by the sports apps.
Besides, 23.73\% of apps do not target a specific type of sport (i.e., the \textit{general} category).
%The results complement the survey of top fan-based sports of the era. 
According to a recent survey of the top fan-based sports by Sports Browser\footnote{https://sportsbrowser.net/most-popular-sports/}, the sports with the most fan bases are in the order of football, cricket, basketball, field hockey, tennis, volleyball, table tennis, baseball, American football, and golf. %And the second and third in our category of sports also in the top ten of the survey. 
6/10 of these most popular sports are reflected in our studied apps in the sports app market while the other four popular sports (field hockey, tennis, volleyball, table tennis) are not. In addition, the number of apps for each sport is not proportional to the popularity of a sport. For example, while golf is the least popular in the top 10 popular sports, it has the second largest number of apps among the top 10 popular sports. The reason might be that users of golf apps are more willing to invest in the apps.
%The missing games are potential market places to launch applications in the future.  \heng{highlight the difference between the number of apps for the sports and the rank from the website; the difference would suggest a imbalance and what need to be improved. Always think about conclusions/take-home messages from the results.} \bhagya{added not sure this is good enough take away message or not}
%However, our dataset or the market trends reflects the fanbase the users of the given game. 
Our results indicate the disproportional distribution of sports apps against the popularity of the sports.

%\heng{We need a table or bullet list to explain the definitions of the functionalities, as some of them are not straightforward, like difference between betting tips and betting, tools, tracking, difference between live updates and news.} \bhagya{updated}
\textbf{Betting tips, training, and live updates are the most popular functionalities provided by sports apps.} The functionalities derived from our manual analysis are defined below:
%\heng{Give an example for each category: For example, the  XXX app provide the YYY functionality to ...} \bhagya{updated}

\begin{itemize}
    \item \textbf{Betting Tips:} This functionality provides users with tips, suggestions and predictions to make bets on a future or ongoing sports competition. For example, \textit{BetsWall Football Betting Tips} provides football, basketball and tennis betting tips based on % with a winning rate of over 84\%. 
    predictions made by artificial intelligence algorithms. %software we have developed in years. %It shares predictions of BetsWall Engine a few hours before the matches start.

    \item \textbf{Training:} This functionality provide physical fitness training for different kinds of sports. For example, \textit{Nike Training Club - Home workouts \& fitness plans (com.nike.ntc)} provides home workouts or mindfulness training for the body and mind preparing for sports. 
    
    \item \textbf{Live Updates:} This functionality provides live updates of the user-selected sports games as push notifications. For example, \textit{BeSoccer - Soccer Live Score (com.resultadosfutbol.mobile)} covers more than 10,000 competitions and enables users to follow games live, and receive notifications of goals, line-ups, news, and signings. % and today’s televised games.
    
    \item \textbf{Streaming:} This allows the users to watch sports games as a video stream. For example, \textit{MyTeams by NBC Sports (com.nbcgeneral.mobile. rsnlocalgeneral)} provides easy access and availability of video sports information to the fans.
    
    %\heng{don't use this kind of ads-like wording. Use your own language to consicely summarize the functionality of the app which is related to the main functionality ``Streaming'' here. Check my changes to the first three and fix this one and the following ones.} \heng{also mind the error signs by Overleaf, for example, you need to use backslash before ``\&''}\bhagya{updated}
    
    \item \textbf{Tools:} This functionality makes use of the sensors in the phone and wearable to help sports players to measure their activities or functions to support sports. For example,  the \textit{YamaTrack Mobile (com.l1inc. yamatrackmarshal)} app provides accurate distances from tee to green and all points in between in a golf course.
    
    \item \textbf{Tracking:} Tracking provides end-to-end data collection and suggestions for sports players to monitor and analyze their activities. For example, \textit{myCloudfitness (com.paradigm.myfitquest3)} tracks all the fitness improvements and gives the necessary suggestions.
    
    \item \textbf{Betting:} This functionality enables users to place bets on competitive sports games. For example, \textit{SuperDraft - a general book Free to Play for Prizes (com.superdraft.superdraftgeneralbookgpf2p)} allows users to place bets on sports events, including Pro and College Football, Pro and College Basketball, Pro and College Baseball, Hockey, MMA and Boxing, Soccer, Tennis. % and more. % or the players of the users choice.
    
    \item \textbf{News:} This functionality provides sports updates and news collected by news reporters. For example, \textit{Onefootball - Soccer News, Scores \& Stats (de.motain.iliga)} gives soccer news, scores, and stats %videos, fixtures, scores, stats, calendar, videos, transfers market 
    of major competitions around the world, including the Premier League, La Liga, Serie A, etc. % Bundesliga, MLS, Liga MLS, Copa Libertadores, Copa Sudamericana and more.

    \item \textbf{Social Network:} This functionality allows different people interested in the same sports to connect and communicate together. For example, \textit{TennisPAL (com.sagedom.tennispal)} provides a platform that helps fans and followers of tennis all around the world to connect and share ideas about the sport.
    
    \item \textbf{League Management:} This functionality helps the users organize their own leagues or keep track and updates of their favourite leagues. For example, \textit{Nizampur Premier League (in.chauka.eventapps.npl)} helps Track all the local cricket leagues and manages the events in Nizampur.
    
    \item \textbf{Radio:} This functionality allows the users to get information about their favourite sports events in the form of audio streams. For example, \textit{Pro Baseball Radio (com.toddbrady.proBaseballRadio)} delivers up-to-date scores, schedules, and standings, and allows users to listen to local radio stations for games live.
    
    \item \textbf{Team Management:} This functionality allows the managers of the sports to organize and track their team's activities in a convenient way. For example, \textit{Soccer Tactics Board (net.soaryong.androidsoccercenter) } helps the managers to track the progress and tactics of the team.
    
    \item \textbf{Health Tips:} This functionality helps the users get specific tips for health and injuries related to specific sports. For example, \textit{Health Diet Foods fitness Help(com.medical.guide\_health.diet.tips)} provides a medical Guide (e.g., Health Diet Tips) specific to fitness programs. %\heng{multiple grammar issues... I could not understand it. Please pay attention to writing}.\bhagya{updated-  sorry for using the description from the application itself. I will modify them }
\end{itemize}
 
Table~\ref{rq1results} shows the distribution of the apps based on their main functionalities. 
Overall, the apps provide a number of similar functionalities.
In particular, betting tips (24.86\%), training (24.86\%), and live updates (15.25\%) are the major functionalities of the apps that we analyzed. % and we can conclude that these three are the emerging market idea for sports apps.  
The results indicate the high competitiveness in these areas and the importance of achieving high quality for an app to succeed, which motivates us to analyze the topics that users are concerned about in their reviews (RQ2) and the aspects that users complain about (RQ3).

%\noindent\textbf

% {\heng{Discussion on the analytics apporaches used in sports apps.}}
% \heng{Treat this as a discussion point at the end of this RQ, as this analysis is weak: we are not sure if an app includes analytic or not without looking into the code. }
%\textbf{ Use of analytical algorithms has been increased in sports apps} 
%\heng{Need table or bullet points to define the analytics methods as they may not be straightforward to readers.} \bhagya{updated}
\textbf{A significant portion of sports apps leverage statistical or predictive analyses.}
%The final categorization we looked into was how much of these apps use any analytical or statistical model. The categorisation analysed are as follows,
The type of analytics derived from our manual analysis are defined below:
%\heng{Give an example for each category: For example, the XXX app use YYY analysis to ...}\bhagya{updated}

\begin{itemize}
    \item \textbf{Predictive analysis:} Some apps use predictive models to make sports-related predictions. For example, \textit{AI football betting tips} is a football betting tips \&  stats project which uses an artificial intelligence-based algorithm. The algorithm calculates all the probabilities and offers betting tips based on past  data, team form, head-to-head results, goals scored, goals conceded, odds fluctuation, team profile (attacking or defending), and past playing formulas.
    %\heng{Please use your own language to concisely summarize the how the example app use predictive analysis, like ``the XXX app use YYY analysis to''... } \heng{fix this one and below}\bhagya{updated}
    
    \item \textbf{Analytical statistics:} Some apps  analyze the data collected and help users interpret the results using predictions for a large group based on a representative sample of the group. For example, \textit{mScorecard - Golf Scorecard} uses statistical analysis, which instantly calculates scores, handicaps, Stableford points  and advanced round statistics. % and distances for up to five players.

    \item \textbf{Descriptive statistics:} Some apps use the data collected to summarise and visualize the data to understand the specific set of observations in them. %\heng{from the definition it is the same as analytical statistics? what's the difference between the two definitions? If there is no difference then the two need to be merged}\bhagya{analytical uses algorithms and analysis whereas descriptive is just plotting values and graphs}. 
    For example, \textit{Peak fitness} uses training plans and progress statistics to manage workout calendars, check clients in for workouts, and track current workouts. 
    
    \item \textbf{Analytical statistics and Predictive analysis:} Some apps use both analytical statistics and predictive analysis. For example,  \textit{Bullet Bet predictions} uses machine learning algorithms (predictive analysis) combined with statistical analysis for more than 200 leagues. It uses machine learning algorithms which analyze historical results, home/away performance and team form to predict the winning team, as well as analytical statistics to understand the distribution against the hypothesis on the larger dataset.
    %\heng{fix this sentence: ``uses artificial intelligence algorithms also with statistical analysis and predictions''; where is predictive analysis used and where is analytical analysis used.}\bhagya{updated} 
    
    \item \textbf{Unknown:} Some apps do not mention the algorithms used but mention that they analyze data. For example, \textit{Free Basketball betting Tips} says it provides Tips but did not mention the strategies for deriving them.
    
    \item \textbf{NA:} Some apps  do not process data of any form.
\end{itemize}

As shown in Table~\ref{rq1results}, we observe that 36.72\% of the analyzed apps use analytical methods. %,\heng{?} the promising 37.85\%\heng{36.72 or 37.85? confusing}. % can be leveraged on further studies on the research areas at the intersection of AI and SE~\citep{Martin2017Survey}. %( Artificial Intelligence for Software Engineering) . 
Sports analytics has gained fast-growing attention in recent years~\cite{morgulev2018sports}.
In our analysis, we observed a significant portion of sports apps (e.g., betting apps) using statistical or predictive analysis to analyze the sports-related data collected, which suggests an interesting research area at the intersection of software engineering and data science in the context of sports apps: integrating data analytics in app development in a responsible way~\cite{clarke2018guidelines}. 
\review{1-2}{In our analysis, a substantial number of sports apps, such as betting apps, were found to employ statistical or predictive analysis to evaluate the collected sports-related data. This observation highlights an intriguing area of research that lies at the intersection of software engineering and data science within the context of sports apps.}
\begin{tcolorbox}
%Key take-homes of this RQ: 1) Sports apps is increasing 2) Diversity of the markets by covering a variety of sport types and providing diverse functionalities 3) Specialty of sports apps: analytics, providing opportunities for future work focusing on analytics in apps
Sports apps have been increasing steadily in recent years, covering a broad spectrum of sports (e.g., football or fitness). In contrast, the distribution of sports apps is disproportional to the popularity of sports. We also observe that the sports app market is highly competitive, with many apps covering similar functionalities. Finally, we observe that a significant portion of sports apps uses statistical or predictive analytics, suggesting an interesting research area at the intersection of software engineering and data science in the context of sports apps.
%Our results motivate us to study the topics that users concern about sports apps (RQ2) and the aspects that users complain about (RQ3).
\end{tcolorbox}
%}

\subsection{RQ2: What are the topics raised by sports apps users in their reviews?} \label{sec:rq2}

\subsubsection{Motivation} 
%\heng{Link motivation the goal of the paper: Prior work (with citations) analyze the topics of general apps in the market. We want to examine if focusing on a specific category (e.g., sports apps) can provide different perspectives. Remove irrelevant text (only keep things that are absolutely necessary).} \bhagya{tried to address them}
% Due to direct access to applications from digital distribution services such as Google Play store, the reviews from the users can be a rich and direct resource of information regarding the application. But how to deal with the plethora of information that is also necessary not in a clean format ( which could contain typos or colloquial usages) can be challenging \citep{al2022designing}. Although, taking leverage of this information source can help in understanding the user experience and product quality, thus defining the application lifecycles to deliver applications with good quality of service and experience.

Previous studies emphasized that understanding user reviews are important, and they contain a lot of information regarding the application behaviour and user impact~\citep{di2017surf,gao2018online,chen2014ar,hadi2020aobtm,di2021investigating}. %\heng{the papers vasa2012preliminary and platzer2011opportunities are from unknown venues, citing them would make your paper look very low. Check papers in the related work section of the paper "On the Importance of Performing App Analysis Within Peer Groups" and cite them instead}\bhagya{updated}. % with NLP challenges to understand \citep{al2022designing}. 
%However there has been many studies on specific category of apps than a generic one which helped to understand the data more better way and get information that can be used to improve the development. 
%Thus we are focusing on sports app study to understand the nuances and collate the information that can give us a different perspective from generic analysis.
These studies usually analyze the topics of general apps in the market. 
In comparison, this RQ aims to examine the user review topics of a specific app category (\textit{sports} apps).
In particular, we want to examine if focusing on such a specific category can provide different perspectives from a general analysis. A study specific to sports industry reviews and their impact has never been analyzed as per our knowledge.

\subsubsection{Approach} 
%\heng{Move the approaches of topic modeling from Experiment Setup to this part. Three sub-approaches: Topic modeling, Selection of the number of topics, Manual labeling of topics. Describe details (e.g., the parameter values) to make it reproducible.}
%\bhagya{addressed}
Our studied 2,058 sports apps have a total of 740,226 reviews. It is difficult to manually examine all these reviews. Thus, we use automatic topic modeling to extract the topics from these reviews.

\noindent\textbf{Topic modeling.}
%Understanding the user reviews is challenging. The challenges in the natural languages, such as the prone tendency to typos and ironies, make the understanding more complicated. 
\review{2-6}{To gain insights into the topics discussed by users, we employed the Gensim wrapper for the Mallet implementation of the LDA topic modeling algorithm\footnote{https://scikit-learn.org/sklearn.decomposition.LatentDirichletAllocation.html}. Mallet  \cite{mccallum2002mallet}, a standalone tool, serves as the underlying framework, while Gensim \cite{gensim-docs} provides a convenient interface for utilizing its functionalities.}
This has been used in many topic modeling-related research~\citep{akef2016mallet,albalawi2020using,o2015analysis}. We performed data cleaning and preprocessing steps as follows. We removed special characters and emojis from the reviews. The reviews we considered are English text only. We created bigram and trigram models using the genism model, followed by the stop words' removal and lemmatization. \review{2-6}{The spaCy model \cite{spacy} is a popular and efficient natural language processing library that provides pre-trained models and tools for various NLP tasks was used for lemmatization.} After preprocessing, a set of words along with their frequencies in each review %dictionary \heng{not a dictionary, should be a list of words for each review?}\bhagya{returned value type is dict, it's words and their freq for the entire corpus} 
%of words was created, which was
were produced and used in the topic modelling. We used the Mallet 2.0.8 version. \review{2-4.5}{In order to mitigate the influence of bot-generated reviews, we implemented a filtering mechanism that excluded reviews containing repeated instances of identical words that exceed a threshold of five occurrences. The threshold is identified with a manual analysis of bot-generated reviews.} 

\noindent\textbf{Selection of the number of topics and other parameters.}
In topic modeling, selecting the number of topics is a critical task. 
We used the coherence score to select the number of topics ($K$). Specifically, we ran our topic modeling using $K$ values from 2 to 50 with a step size of 1 and chose the $K$ value with the highest coherence score. 
%We iterated several topics from 2 to 50, calculated each topic's coherence value, and chose the topic with the best coherence score. 
%The evaluation metric that we used to select our topics is a coherent score, and we run out topic models from topic number 2 to 50 with a step size of 1. 
Out of all the 49 models, we selected the model with a $K$ value of 17, which achieved the highest coherent score of 0.63. %0.6288.
\review{2-10}{A higher coherence score indicates that the words within a topic are more closely related and that the topics are more well-separated from one another. Existing studies show that such a coherence score is reliable when it is greater than 0.5 \cite{syed2017full}. The choice of the coherence score threshold for choosing the number of topics (k) can vary depending on the dataset, research context. We used the an elbow point or a point of diminishing returns in the coherence score plot to determine a reasonable range of k values.}
Besides, other parameters we used in the topic modelling include the optimization interval of ten, the number of iterations of 1000, the alpha value of 5.0, and the beta value of 0.01\footnote{https://senderle.github.io/topic-modeling-tool/documentation/2018/09/27/optional-settings.html}. %\heng{double check if alpha value is correct}. \heng{how about beta?}\bhagya{updated} 

\noindent\textbf{Analysis of topics.}
Once we had the final topics, we manually labeled each topic. To label each topic, we looked into the representative words and the reviews that contributed most to that topic. During the manual labeling of topics, we merged topics that are highly similar to each other \cite{stackoverflow}. %Thus, we could analyze the point of discussion about the apps discussed. 
In particular, we are interested in how the resulting topics are different from the topics derived from the reviews of general apps.
We also examined
%We also looked into how it differs from the general topics of reviews for other apps and
how the topics are distributed across different categories of sports apps in terms of the main functionalities. 
%The following categories were the analysis: Dominant topics in each review, Representative review for each topic, and Distribution of topics across reviews.
The labels defined were reviewed by four mobile app developers from a Google developer group (GDG Cochin). %\heng{are they the same as the Google developers? be consistent in wording}\bhagya{addressed}, 
%we connected with the 4 people from GDG Cochin to participate the blind review.
To validate the grouped topics for each review, we considered the topic percentage contribution metric. For each review, they have their topic contribution metric for the classified topic indicating the likelihood that a topic is represented in the text. We capped a threshold of 0.075 from the graph distribution of topic percentage contribution against the number of reviews. Value 0.075 covers 80\% of the total number of reviews. 
%\heng{explain why using a threshold and why this specific value}\bhagya{updated} 
The topics with the given topic percentage contribution or above are considered to be valid topics for the classified review.
\subsubsection{Results} 

%The evaluation metric that we used to select our topics is a coherent score, and we run out topic models from topic number 2 to 50 with a step size of 1. Out of all the 49 models we selected the model with topic number 17 which has the highest coherent score of 0.6288. 
%We evaluated the model with 17 topics and manually labelled the topic categories based on the contributing words and most contributing reviews. 

\textbf{We derived 14 topics from the reviews of the sports apps}.
Our derived topics, along with their most significant words and the assigned labels, are listed in Table~\ref{tab:topic-summary}.
\textit{Positive feedback} %\heng{use a different font like italic for topics or other special noun phrases} 
is the dominant topic (19.27\% in total reviews and 32.50\% in the sample set) of the user reviews and it is the most common topic. %Also when we analyse the distribution of topic generated over the corpus positive feedback, \heng{use consistent upper/lower cases}
In addition, \textit{Up-to-dateness} (10.15\%) and \textit{Quality of Content} (10.67\%) are among the frequently discussed topics, whereas \textit{Bug Reports} (3.91\%), \textit{Negative Feedback} (3.85\%) and \textit{Advertisement} (3.45\%) are the least discussed topics, as shown in Table~\ref{tab:topic-summary}. % and Fig \ref{topic-func-distrib}\bhagya{should we remove this fig?}. 
In the following section, we discuss each of the 14 topics and why they are significant for users and developers. In Table~\ref{tab:topic-summary}, we have added cross-references to the previous works where a similar topic has been discovered in previous works on app review analysis. %both general app analysis and specific app analysis. %\textbf{3 new topics identified on sports app based study}

\begin{table*}[t!]
    \centering
    
    \caption{Topics extracted from the reviews of sports apps. %\heng{sorted the topics by their percentages. For future records, always sort information by their importance if appropriate}\bhagya{added to my notes thank you}
    %\heng{Some times like ``Positive Feedback'' other terms like ``Streaming quality'': use consistent upper/lower cases.}\bhagya{updated}
    } 
    % \heng{can we add the percentage of reviews for each topic here? It's different from Table 3 which only includes the smaller manually analyzed apps} \heng{explain what the star symbols mean in the table note or merge the same labelled topics together (two rows of keywords separated by a horizontal bar in the middle).} %\bhagya{the topic names are very different from each other in all the reference papers however, it is the same meaning so kept the actual references to themselves.}
    %\bhagya{however, the topid review credibility mapped back to the previous research work and removed from the new finding topics}
    %\large
    % \caption{Summary of Topics generated. \heng{Replace the column topic percentage contribution with the percentage of reviews with that topic as the dominant topic (the percentages should add to 100\%), and sort the topics by the distribution. Add a new column to provide an example review. Add another new column to indicate if a similar topic is identified in previous work (with references) or it's a new topic, similar to the last column in Table 4 in paper: https://www.hengli.org/pdf/Li2020LoggingQualitativeStudy.pdf}}
    \label{tab:topic-summary}
    \scalebox{0.75}{

    % \tiny
    \begin{threeparttable}
    \begin{tabular}{c c c r c}\toprule

    \textbf{\specialcell{Topic\\Number}}&	\textbf{\specialcell{Topic\\Label}}&	\textbf{Key Words}&\textbf{\specialcell{Total\\Con.}}&	\textbf{\specialcell{Prior\\work}}\\\midrule
   
     \multirow{5}{*}{1} & \multirow{5}{*}{\specialcell{Positive \\feedback\tnote{*}}}	&	\specialcell{great, easy, love, helpful, follow,\\ app, simple, fun, super, fast} & \multirow{5}{*}{19.67\%} & \multirow{5}{*}{\specialcell{\citep{gu2015parts},\\\citep{zevcevic2021user}}}	\\

   &	&	\specialcell{good, app, nice, world, job,\\ graphic, usefull, enjoy, awsome, pretty}&  &	\\
    
    &	&	\specialcell{app, good, great, awesome, excellent, \\cool, experience, enjoy, site, woman}& \\\midrule
    
    \multirow{3}{*}{2}&	\multirow{3}{*}{\specialcell{Quality of\\content\tnote{*}}}&	\specialcell{love, amazing, perfect, fitness, recommend,\\enjoy, daily, yoga, absolutely, class}& \multirow{3}{*}{10.67\%} & \multirow{3}{*}{\citep{zevcevic2021user}} \\
    
    &	&	\specialcell{workout, exercise, plan, home, gym,\\set, training, routine, choose, rest}& \\\midrule
    
    3&	\specialcell{Up-to-\\dateness}&	\specialcell{game, team, match, play, score,\\ football, live, news, player, fan}&10.15\%& \textbf{New}	\\\midrule
    
    4&	\specialcell{Return on \\investment}&	\specialcell{free, pay, money, subscription, month, \\version, year, worth, buy, cancel}&8.86\%&\specialcell{\citep{kalaichelavan2020people},\citep{mcilroy2016analyzing}\\\citep{cen2014user},\\\citep{zevcevic2021user}}	\\\midrule

    5&	\specialcell{Version\\update}&	\specialcell{work, update, phone, android, screen,\\ fine, version, late, slow, device}&7.62\%&	\citep{Martin2017Survey},\citep{mcilroy2016analyzing}\\\midrule

    6&	\specialcell{Streaming\\quality}& 	\specialcell{watch, tv, show, live, stream,\\ channel, service, video, minute, freeze}&6.91\%& \citep{Martin2017Survey}\\\midrule

    7&	\specialcell{Impact of\\application}&	\specialcell{day, feel, lose, week, weight,\\start, body, result, exercise, fit}&6.54\%&	\citep{zevcevic2021user}\\\midrule

    8&	\specialcell{Feature\\request}&	\specialcell{option, video, notification, user, put,\\ view, find, add, read, turn}&5.37\%&\specialcell{\citep{villarroel2016release},\\\citep{kalaichelavan2020people},\citep{gu2015parts}\\\citep{mcilroy2016analyzing}}	\\\midrule

    9&	\specialcell{Accuracy of\\prediction}&	\specialcell{application, tip, bet, prediction, guy,\\ win, amazing, good, accurate, vip}&5.27\%&	\textbf{New}\\\midrule

    10&	\specialcell{Review\\credibility}&	\specialcell{app, give, download, star, sport,\\ information, review, rate, full, info}&4.30\%& \citep{di2017surf}	\\\midrule

    11&	\specialcell{Bug\\reports}&	\specialcell{fix, problem, issue, open, log,\\ load, crash, sign, account, email}&3.91\%&\specialcell{\citep{villarroel2016release},\\\citep{gu2015parts},\\\citep{cen2014user}}	\\ \midrule
    
    12&	\specialcell{Negative\\feedback}&	\specialcell{make, worst, app, find, change,\\thing, issues, program, people, hate}&3.85\%&\citep{fu2013people}\\\midrule

    13&	\specialcell{Advertisements}&	\specialcell{time, ad, bad, back, start,\\long, waste, close, stop, annoying}&3.45\%&\specialcell{\citep{kalaichelavan2020people},\\\citep{cen2014user}}\\
    
    \midrule

    14&	\specialcell{Precision of\\tools}&	\specialcell{add, track, feature, run, show,\\datum, step, point, save, record}&3.43\%& \textbf{New}\\

\bottomrule
    
    \end{tabular}
    \begin{tablenotes}
    \item[*] The topic is merged from multiple LDA-produced topics in our manual labelling. Therefore, there are multiple sets of keywords.
    \end{tablenotes}
    \end{threeparttable}
    }

\end{table*}

%\heng{Discuss the topics by two groups: 1) topics that are identified for apps in general (similar topics are identified in prior work); 2) topics are are unique to sports apps (new topics). The latter is the highlight and contribution of the work.}
%\bhagya{I haven't differently classify it to two groups but mentioned in the description that they are new findings - the categorisation is done in the table}
%\nd

%\heng{updated the order}

1) \textbf{Positive feedback:} It is the most popular topic in user reviews. The users express their general positive feedback to an app in such reviews. Below is an example:
%The feedback system for the upcoming users. They explain how they feel about the apps are great, love, simple, absolutely, brilliant, fan etc .
\begin{leftbar} \noindent
\textit{ ``I love this app very much."} %\heng{is it possible to use a smaller gap between the bar and the text?}\bhagya{updated}
\end{leftbar}

2)\textbf{Quality of content}: %Reviews are a way to convey how users feel about the apps. And as we mentioned in our previous section, surveys show 79\% of people tend to check the reviews before downloading. For example, this review 
Many sports apps deliver sports-related content (e.g., instructions, videos, news, or radio broadcasting) to users. The quality of such content is of critical importance to the users. 
\begin{leftbar} \noindent
\textit{``Even if you barely have time to work out this app will work for you Easy to understand, clean and clear instructional videos...''} %\heng{no need to keep the irrelevant part, remove the rest of the review. Similar rule applies to other topic examples as well }\bhagya{Noted}, the interface is nice One of the best workout apps "
\end{leftbar} 
%, indicate the ease of use of the app, and its interface.

3) \textbf{Up-to-dateness}: %As we observed in RQ1, live updates is one of the main functionality of the sports apps. Being the purpose the rather than the delivery the accuracy and timeliness is also concerned metric for the users which can clearly visible in the reviews like 
Many types of sports apps need to provide regular updates to the delivered content to the users, such as apps that provide betting tips, live updates, or news. The timeliness of such updates is one of the most concerning topics for the users of these apps. 
\textbf{This is a new review topic that has not been previously identified in other studies of app reviews.}

\begin{leftbar} \noindent
% \textit{``The push notifications didn"t have what happened in the game, just the score. And there"s no box score or link to one, to find out what happened. If my team scored a run, I don"t know who scored it, who knocked him in, who moved on the bases, etc.}
\textit{``This app helps me get notifications of scores and when matches are starting because if I don't have WiFi, they will just send me I'd somebody scores or I'd a match started.''}
%\heng{This review is not related to timeliness/speed. Is there a better review? or is the labelling ``timeliness of updates'' correct, or ``update-to-dateness'' (still need a better example)?}
% \bhagya{updated the review - can discuss the title }\heng{I'd use ``update-to-dateness''}\bhagya{updated}
\end{leftbar} 
%This metric that we concluded was also unique from previous topic modeling models, where it the quality of the apps are measured with the element of time too.

4) \textbf{Return on investment}: %From our basic analysis of our dataset, we understood that 
24\% of the sports apps in our dataset are paid and developed as a business. %Thus the users are curious to evaluate 
%The app's value or performance based on the money they invest in it.
Even when an app is free, the users usually need to invest time in the advertisement.
The users are concerned about the value or performance of the app based on the money or time they invest in it. 
\begin{leftbar} \noindent
\textit{``It's designed to take as much of your hard-earned cash as possible. Constantly need energy drinks, and your salary just doesn't keep up with the cost if you pay for the full ad-free version, then this shouldn't be as bad as if playing for free."}
\end{leftbar} 

5) \textbf{Version update}: This topic mainly discusses the challenges the users face with version updates or the differences between the two versions, compatibility of the version with devices and use case. And this will help us, the developers, to decide the impacts of each version and mitigate and resolve issues associated with them. %\heng{not clear how it is related to release patterns}\bhagya{my idea was when there is an issue with one version - the next version should be released faster, thus it will help to arrange the release patters}. 
\begin{leftbar} \noindent
\textit{``update won't install on note 9''} %\heng{why there are missing `'' in many of these reviews. If it is a result of removing punctuation in the processing step, please use the original reviews or fix these sentences. Applied to other review examples as well}
%\bhagya{most of the reviews extracted are not complete sentences, these are the raw reviews itself, changed the review}
%\heng{this review is not about ``update'', it's about difference between different versions. Use a better example or improve the label (such as ``versioning''?).}
\end{leftbar} 

6) \textbf{Streaming quality}: Watching sports videos live is vital for sports fans. The quality of the streaming bothers them and thus becomes the topic in their reviews. Audio broadcasting through radio to discuss sports reviews also contributes to the given topic.
\begin{leftbar} \noindent
\textit{``The app needs work. This review is for the app and not the fight pass service. There is no option for offline viewing, so every video is a stream, which this app sucks at. You can't fast-forward without the playback freezing and having to start over. Live streams that I've watched were VERY problematic, and important exchanges were missed due to freezing. If I'm expected to pay this much for a service this frustrating app can't deliver, what's the point? "} 
% \heng{this review is not about video streaming quality, it's about an UI design, please find a better example like searching the keywords freeze} \heng{Again, please add the removed punctuations back to the reviews, otherwise the sentences are not readable}
\end{leftbar} 

7) \textbf{Impact of application}: 
%The main difference between positive feedback and impact was that, in impact the users are evaluating the impact that the apps have on the society rather that individual level. 
Many sports apps (e.g., training or tracking apps) can have certain physical or mental impacts (e.g., improving fitness) on the users. The impacts of such apps are an important topic raised in users' reviews.
\begin{leftbar} \noindent
\textit{``Lose Weight and Burn Belly Fat with fat burning workouts for women at home. The best lose weight app for women to burn fat and lose weight at home! With simple and effective fat-burning workouts for women, you can lose belly fat, lose thigh and arm fat. Follow the 30-day plan and take just a few minutes a day to lose weight and get in better shape! 2-7 min fast workouts and HIIT workouts allow you to lose weight and keep fit anytime, anywhere. No excuse anymore! You can track burned calories"} 
% \heng{Unreadable, please use the original review without any pre-processing}
% \heng{not sure how it is about impact of application, maybe better find some review related to impact such as losing weight?}
\end{leftbar} 

8) \textbf{Feature request}: Users not only express opinions on the existing features but also tend to convey what they like through their reviews. User reviews on this topic itself will give companies a plethora of ideas to improve their apps. Prior studies~\citep{villarroel2016release,kalaichelavan2020people,gu2015parts} have proposed approaches to mine feature requests from user reviews.
\begin{leftbar} \noindent
\textit{``Awesome app for beginner, But need some offline video feature otherwise this is great."} 
% \heng{missing punctuation, fix everywhere}
\end{leftbar}  

9) \textbf{Accuracy of prediction}: From our analysis in RQ1, we understood that over 22\% of apps (e.g., betting tip apps) use some form of predictive analysis.  For those apps, the accuracy of the predictive analysis is an important topic raised by users in their reviews. %essential and the top market from our analysis was betting it needs more accurate results and the users are concerned about the accuracy of the systems used. 
\textbf{This is a new review topic identified in our work}. Although the topic name ``accuracy'' has been mentioned in works like \citep{fu2013people}, however, the context of using accuracy in our work is mostly on the accuracy of the predictions in the concerned apps. %and recall of the machine learning and analytics models used in the application.

\begin{leftbar} \noindent
\textit{``Honestly, a correct prediction is almost impossible bcz you can't control what goes on in the field of play, but this app has been amazing..I give u 5 stars. good job guys!!"} 
% \heng{this seems not an user review but an ads}
\end{leftbar}

10) \textbf{Review credibility}: %This is an interesting topic that has discovered. 
In the reviews on this topic, the users validate the previous reviews with their own personal experiences.
\begin{leftbar} \noindent
\textit{``Can't trust reviews. All of the reviews are five stars and tons of them are all posted on the same day as if it's the employees or paid reviews. Just suspicious. I downloaded it and immediately can't even see if I want the app before it asks for you to pay for a year. I then really check out the reviews and they look fake. After every semi-negative review, they flood the app with tons of positive ones. So SUS"}
\end{leftbar} 
%From the further analysis we understood that it is a unique topic that we identified which never been mentioned in the previous works. 

11) \textbf{Bug reports}: While testing apps, there are always constraints about testing all the scenarios. But the users are the best testers out there with a different setup. They will explain the challenges to using the apps, and those can be resolved for the better performance of the apps. 
\begin{leftbar} \noindent
\textit{``When system errors occur, app doesn't save your place. The user has to start over and if it's been more than a few minutes into the workout, the app finishes the workout, which is more hassle to redo it. "} 

% \heng{fix punctuation}
\end{leftbar} 

12) \textbf{Negative feedback}: In contrary to the ``positive feedback'' topic, users express their general negative feedback to an app in reviews on this topic. They typically use short messages to express their frustration like the one below. 
%Critiques are the second most expected topic, and there are plenty of short messages with their frustration filled.
\begin{leftbar} \noindent
\textit{``Most disappointing app I have used in my life"} 
\end{leftbar} 

13) \textbf{Advertisement}: From our previous analysis, we understood that the majority of the apps are free, and the profit of the apps generally comes from the ads in the apps. However, the ads have a huge impact on the user experience which is discussed by the users in their reviews. 
\begin{leftbar} \noindent
\textit{``Would have given it a 4star but come on, what is it with these ``ads?" Ads keeps popping up even if I close it, the only way out is to ``force stop" the app, delete data and run the app again. This happened to me last year until an update put everything back to normal, now it's started again, showing Guinness ads over and over again, I don't drink Guinness for God sake.."}
\end{leftbar} 

14) \textbf{Precision of tools}: As discussed in RQ1, there are mobile apps which take leverage of the sensors in mobile phones or wearables to measure distance, speed etc. and act as a tool to support the sportsmen. The precision of the tools is vital for the success of the application and thus an important concern of the users as expressed in their reviews. As the sensors of different mobile devices may have different accuracy, it is important that app developers test their apps on mobile devices of diverse sensing capabilities. 
%The quality of apps using the sensors are very much dependent on the precision of measurement that it provides. 
%The specific study on the sports apps uncovered the topic on the importance of sensor precision in the app success thus on the user experience.
\begin{leftbar} \noindent
\textit{``Tracks step count during a run but not cadence (steps/min)Spoken announcements only at 0.5km and 1km; no 400m option or ability to enter a custom distance graphs are tiny and coarse like Samsung Health but at least the plot points here are selectable. Missed the opportunity to match plot points with points on map like in Nike"s running app No auto-pause, not even a simple one let alone a good one like Strava"s No split or lap times All in all, very basic as a running app and can"t stand alone."}
\end{leftbar} 
\begin{figure}[!t]%{0.49\textwidth}
         \centering
         \includegraphics[width=.7\textwidth]{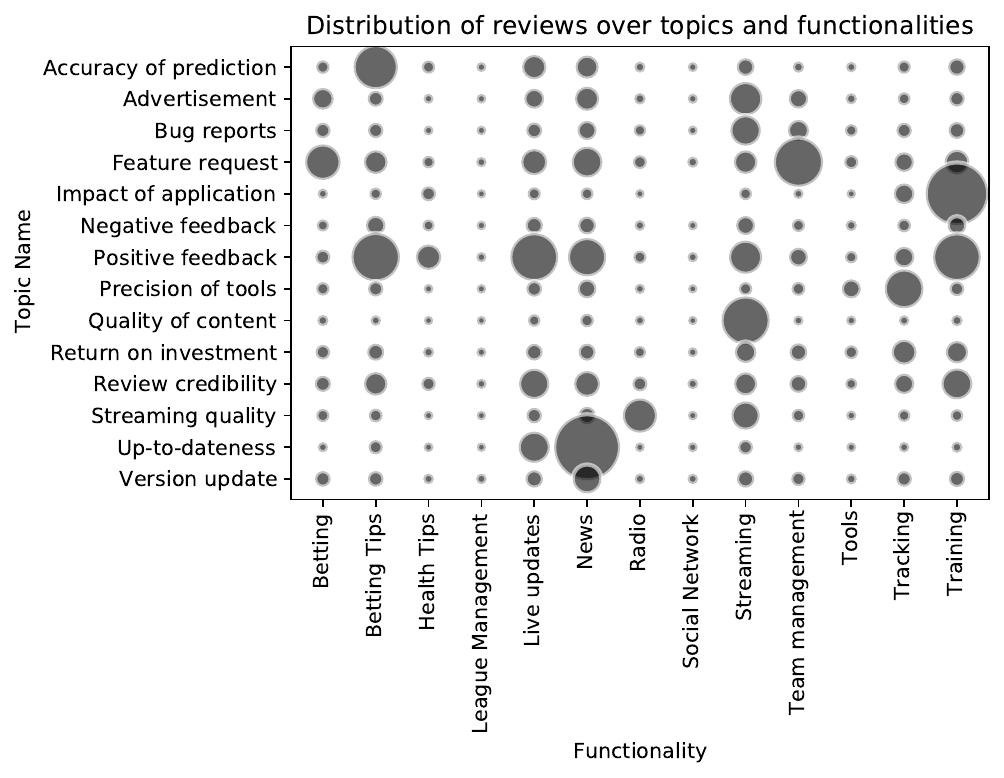}
         \caption{Distribution of reviews over different functionalities and topics. %\heng{Great! Makes a lot of sense!  Use ``Up-to-dateness''}\bhagya{updated}
         }
         \label{fig:functions-to-topics-distibution}
     \end{figure}
\textbf{While most of these topics are found in previous review analysis studies, we identify new topics that are special for sports apps, including \textit{accuracy of prediction}, \textit{up-to-dateness}, and \textit{precision of tools}}.
Prior works have analyzed app reviews to identify the topics in users' reviews. They either analyze apps across different app categories (e.g.,~\cite{chen2021should,hassan2022importance,gao2018online}) or focus on a small set of particular apps (e.g.,~\cite{frie2017insights,brown2020review}). 
In comparison, this work analyzes the review topics of a large number of apps in a single app category (\textit{sports} apps). 
We observed that, out of the 14 topics derived from our analysis, 11 topics could be mapped to similar topics identified in previous works. For example, we recognized topics such as the impact of application which has been the focus of study on its own~\cite{frie2017insights,brown2020review}. However, our analysis of the review topics of such a single app category can provide more specific insights into the app category, as we identified three review topics (\textit{accuracy of prediction}, \textit{up-to-dateness}, and \textit{precision of tools}) that are associated with the characteristics of the sports apps. For example, as many sports apps (e.g., betting tip apps, live update apps, etc.) need to provide regular updates to the delivered content, the up-to-dateness of the content of the apps is one of the most concerned topics of the users. \review{2-11}{Our results suggest that future work on app review analysis can pay more attention to analyzing many apps in individual app categories. For example, considering the fact that a single category itself contributes to a fiercely competitive market with a large number of apps, a category-based study helped to identify the distinctive solution to the bugs and set the standard amongst the competitors. For example, a unique concern (e.g., up-to-dateness) may be raised by users of a particplar category of apps (e.g, sports apps). Besides, topic prevalence and sentiment (as discussed in RQ3) help to prioritize the issues specifically among the competitors to improve user retention.
}

%\bhagya{should we add a paragraph on topic function distribution with a figure or this is enough??}

\textbf{The distributions of review topics vary across apps with different main functionalities.}
Fig \ref{fig:functions-to-topics-distibution} shows the analysis of the reviews over different functionalities and different topics. As we concluded previously, the \textit{negative feedback} bubbles overall functionalities are relatively small, whereas \textit{positive feedback} is bigger, thus the users are generally happy about sports apps and their use cases. The topic \textit{accuracy of prediction} has a significantly higher distribution for apps with the main functionality of \textit{betting tips} which indicates the importance of using more accurate machine learning for these apps. Similarly, topic \textit{precision of tools} is highly cluttered in the functionality \textit{tracking} which uses support/information from build sensors of phones or wearable devices, % (i.e., apps with the main functionality \textit{tracking} 
which indicates the necessity of adaptability of applications to the cutting-edge technologies of wearable devices or general IOTs. The topic \textit{up-to-dateness} is mainly associated with two functionalities \textit{news} and \textit{live updates}, which deals with the timeliness delivery of the information to the users, which indicates the importance of keeping the information updated and push notifications in these mobile applications.
\begin{tcolorbox}
%Our focus study on sports app categories helped to unwind 4 new topics such as review credibility, accuracy of prediction, update-to-dateness, and precision of tools that have not been discussed in previous studies. Also studies specific to single categories than generic ones helps to understand the user reviews better with more intuitive results.
We derived 14 topics from the user reviews of 2,058 sports apps. 
While most of these topics are found in previous review analysis studies; we identify new topics that are special for sports apps, including \textit{accuracy of prediction}, \textit{up-to-dateness}, and \textit{precision of tools}.
Our results suggest that future work on app review analysis can pay more attention to analyzing a large number of apps in individual app categories.
\end{tcolorbox}
%}

\subsection{RQ3: What do users complain about sports apps?} \label{sec:rq3}

\subsubsection{Motivation}
%Sentiment and Irony are one of the biggest challenges in natural language processing. 
%Understanding whether the users are ironic about the reviews is crucial while understanding the review. 
%The success of the application depends on the good reviews and the bad reviews are the methods to improve the applications. 
%A previous survey by industries\footnote{https://tapadoo.com/} shows that users rely on the previous reviews to authenticate the users. 
App users express their feedback about an app in their reviews. Understanding what users complain about can provide insights for app developers to improve their apps' quality or meet the users' needs~\cite{fu2013people,kalaichelavan2020people,mcilroy2016analyzing}. 
%Investigating the user complaint factors and mitigating them will help improve the app's quality. 
%And sentiment analysis is how to unravel the users' challenges expressed via text reviews. 
In RQ2, we identified the topics that users of sports apps discuss in their reviews. In this RQ, we further analyze which factors (topics) users complain about, accounting for both the general topics (e.g., quality of content) and the topics specific to sports apps (e.g., up-to-dateness).

\subsubsection{Approach}
%\heng{Organize the approach section like the other two RQs. Move the details of sentiment analysis from Section 2 to this RQ-Approach.}\bhagya{updated}
%Finding the reason the users are complaining most about the sports apps.
%Although we analyzed the topics discussed in the reviews in our RQ2, this section examines the next level of the details to understand why users hate sports apps.
In this RQ, we analyzed the existing user review analysis method to understand why users complain about sports apps.
The previous study~\cite{fu2013people} has been done on the generic apps to understand what users complain about the apps~\cite{fu2013people}. Each review is associated with a rating, the quantifiable measure of the users' feedback about an app. The review is the descriptive way of the same feedback. Hypothetically, both are significantly correlated: the reviews with low ratings should be bad, and those with high ratings should be good. 
%However, the cases are different. 
However, there is a lot of noise in Text-Rating-Consistency(TRC)~\cite{fu2013people}. %, as explained in the paper. 
The approach we used to mitigate this challenge is to evaluate the sentiment of the reviews and correlate it with the review ratings available. %And we removed the inconsistent outliers.

The intuition is that negative sentiment is associated with negative feedback in the user review. 
Prior work has used a similar sentiment analysis of user reviews to understand how users like the features of an app~\cite{6912257}.

% Understanding and making the sentiment of users' emotions while writing the reviews can aid in understanding what they are conveying through the reviews. 

\noindent\textbf{Sentiment analysis.} 
We use NLTK’s pre-trained sentimental analysis library called VADER (Valence Aware Dictionary and sEntiment Reasoner)\footnote{\url{https://github.com/cjhutto/vaderSentiment}}. VADER appears to be the best suit for the language used in social media~\citep{hutto2014vader}, which are short sentences with some slang and abbreviations expected to be similar in user reviews. The median review length of the entire review corpus under study is 12 words per review sentence. 
\review{3-2}{For each review ( Google reviews have a max limitation of 500 characters) that we have in our dataset, VADER analyses the sentences and returns four different scores for each review: a negativity score (neg), a neutrality score (neu), a positivity score (pos), and a compound score (compound). The compound score is a normalized, weighted score that combines the three sentiment scores (neg, neu, and pos) to produce an overall sentiment score ranging from -1 (most negative) to +1 (most positive). VADER is sensitive to both Polarity (whether the sentiment is positive or negative) and Intensity (how positive or negative is sentiment) of emotions. When a sentence has both positive sentiment and negative sentiment in it, either the positive polarity or negative polarity will be higher than the other. For example a sentence like ”The app looks nice but is complicated” has polarities of ’neg’: 0.0, ’neu’: 0.556, ’pos’: 0.444, ’compound’: 0.3182, whereas for the sentence “The app look nice and The app is complicated” the polarities are  ’neg’: 0.433, ’neu’: 0.567, ’pos’: 0.0, ’compound’: -0.2732. From the above example itself it is clear that the compound polarity is a useful metric that can account for the presence of ironies and contrasting sentiments in a sentence. Moreover, VADER relies on a dictionary that maps words and other numerous lexical features common to sentiment expression in microblogs. These features include: A full list of Western-style emoticons ( for example - :D and :P ), Sentiment-related acronyms ( for example - LOL and ROFL ), Commonly used slang with sentiment value ( for example - Nah and meh )}
 %As we focus on reviews with negative sentiment (i.e., what users complain about), 
Following the guideline\footnote{\url{https://github.com/cjhutto/vaderSentiment}} by VADER, we use the \textit{compound} sentiment score to classify the sentiment: positive sentiment (score $>=$ 0.05), neutral sentiment (0.05 $>$ score $>$ -0.05), and negative sentiment (score $<=$ -0.05), % we capped the threshold from -0.05 to -1 as the negative sentiment
similar to prior work~\citep{elbagir2019twitter}. Although different adaptive ranges of sentiments were selected for the reviews based on the distribution by ~\citep{borg2020using,hutto2014vader}, we are sticking to the basic conventions.  %\heng{Are these two citations used to support the range of -0.05 to 1; please put them at the right place }\bhagya{these are the citations examples of using different ranges other than the one we used}
We also experimented with sentiment analysis tools such as the Textblob library\footnote{https://textblob.readthedocs.io/en/dev/}. Our results show that VADER gives better results which confirm others’ observations\footnote{https://www.analyticsvidhya.com/blog/2021/10/sentiment-analysis-with-textblob-and-vader/}. 

\noindent\textbf{Correlation analysis.} 
%To isolate the complained factors from the users in a very imbalanced data set, we used the following metrics ( Fig ~\ref{correlation+}).
To understand the relationship between a user's sentiment in a review and the rating given to the reviews by the users, we perform a pairwise Spearman correlation analysis between the following metrics:

\begin{itemize}
    \item \textbf{Average user rating:} Arithmetic mean of the ratings associated with all reviews of sports apps for each app.
    \item \textbf{Average sentiment:} Arithmetic mean of sentiment calculated for all reviews for each app.
    
    \item \textbf{Percentage negative rating:} Distribution of negative rating ($<=$ 2) to the number of all reviews associated with each app.
    \item \textbf{Percentage negative sentiment:}  Distribution of negative rating ($<=$ -0.05) to the number of all reviews associated with each app.
    
\end{itemize}

For each pair of metrics, we calculate Spearman's correlation coefficient and the p-value~\cite{akoglu2018user}. The results are shown in Fig.~\ref{correlation+}.

%a) \textbf{Preprocessing}: 
\noindent\textbf{Removing outliers.}
User reviews are one of the pure forms of natural language; it contains a lot of noise, and the tricky part of natural language - is the irony. \review{3-7}{User reviews are one of the forms of natural language; it contains a lot of noise, and the tricky part of natural language - is the irony.  Irony, as a linguistic phenomenon, involves the use of words or expressions to convey a meaning that is often contrary to the literal interpretation. For example consider the given review "Wow, this fitness app is truly revolutionary! I love how it constantly reminds me of how out of shape I am and how many calories I've consumed. It's just what I needed to boost my self-esteem and make me feel like a fitness superstar. The daily notifications and relentless tracking make me feel like I have a personal trainer who never leaves my side. Who needs motivation and positive reinforcement when you can have an app that ruthlessly points out your flaws? Thanks, fitness app, for reminding me why I never liked working out in the first place!". In this review, the irony is evident in the sarcastic tone used to describe the fitness app. The reviewer humorously emphasizes the app's features that may not be traditionally considered as motivating or positive, such as constant reminders of being out of shape and tracking calories. By highlighting the app's relentless approach and lack of encouragement, the reviewer pokes fun at the idea that such features would genuinely inspire and uplift users. This complex form of communication can be difficult to detect and accurately analyze using computational methods \cite{weitzel2016comprehension}. In particular, the inherent ambiguity and subtlety of ironic statements make it challenging for rule-based models or algorithms to accurately identify and interpret such instances\cite{weitzel2016comprehension, tymann2019gervader}.} %We used the NLTK sentimental analysis library to return each review's positive, negative, and compound values. 
\review{3-2}{In order to detect and eliminate outliers, we utilized both the sentiment score and user rating of each review. Specifically, we examined instances where the rating was less than or equal to 2, indicating a negative review, and where the sentiment score was not less than -0.05. In such cases, we considered the review as an outlier and excluded it from further analysis~\cite{hassan2022importance,mcilroy2016analyzing,hassan2018studying,noei2019too}. Similarly, we identified instances where the rating was more than or equal to 4, indicating a positive review, and where the sentiment score was not more than 0.05, and treated them as outliers as well.}

%\heng{Add a paragraph for the approach of analyzing complained factors by app functionality}
%\bhagya{added at the end}

\noindent\textbf{Functionality level analysis.}
We further analyze the user-complained factors for each group of apps with the same main functionalities, we indicated it as functionality level analysis.
Understanding the complained factors on a functionality level gave more intricate information about the concerns users have for a specific group of apps. Functionality level analysis is a comparison of the topics generated for each review to the functionality of the sports application with which the reviews are associated with. 
For each group of apps with the same main functionality (from the manually studied sample in RQ1, see Sec~\ref{sec:sample}), we analyzed the %sample set (Sec~\ref{sec:sample}) with 
average sentiment of the reviews and the percentage of negative sentiment reviews associated with each topic. % and each function.

%\heng{what did you do to the outliers, remove them? I think the outliers might be a good discussion point.}. \bhagya{we addressed this as the outliers but not in detail why it happened}

% \heng{The rest of the approach may be removed unless the previous parts miss anything}

% \newline
% b) \textbf{Sentiment analysis}: We analyzed the sentiment of the reviews and plotted against the user reviews to understand the users' emotions while writing the review.

% \vspace{8pt}
% \noindent\textbf{The factors users most complain about}

% \begin{itemize}
%     \item  \textbf{Data collection}: We have the bad reviews without the outliers and the topics associated with each review from our previous sections. Also, we considered these given research papers to understand the common hate factors among the available apps and another category of apps to list out the significant points to the differences in the disturbing characteristics for the users.
%     \item \textbf{Analysis}: We compared the topics that are mainly being discussed in the low rated reviews. To understand more on reviews of complaints we did Qualitative analysis why these topics are associated with negative sentiment and low ratings: use a stratified sampling of the curated data set  with 95\% confidence level and 5\% confidence interval.

% \end{itemize}

\subsubsection{Result}
\begin{figure}[htbp]
\centering
\includegraphics[width=0.85\textwidth]{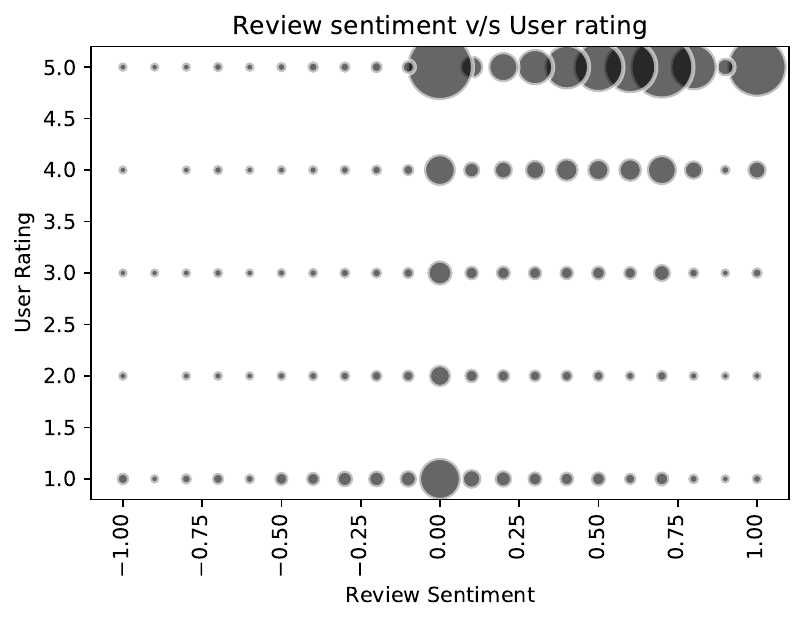}
\caption{Distribution of review sentiment over user rating associated with reviews (before reviewing the outliers). %\heng{this is before removing the outliers, right?}\bhagya{sorry! yes, this graph is before outliers, should we keep this or the one after removing outliers? }
}
\label{rating}
\end{figure}

\textbf{Users are generally happy about the sports application and their performance.}
Figure \ref{rating} describes the distribution of reviews over the review sentiment values and the user ratings. 
Overall, 94.44\% of the reviews have a positive sentiment (i.e., sentiment score $>=$ 0.05), while only 5.56\% of the reviews have a negative sentiment (i.e., sentiment score $<=$ -0.05).
65.5\% of the reviews with negative sentiment are associated with negative ratings (e.g., rating $<=$ 2).
In comparison, only 34.5\% of the reviews with positive sentiment are associated with negative ratings.
As we can see, there are a few outlier instances with very negative sentiment reviews (e.g., ``Not working", with a sentiment of -0.9) but with a very high (e.g., a rating of 5). To mitigate the text rating inconsistency, we removed the outliers to make the analysis fairer, as described in the approach section.

\begin{figure}
    \centering
    
    \begin{subfigure}[t]{0.75\textwidth}
        \centering
        \includegraphics[width=\textwidth]{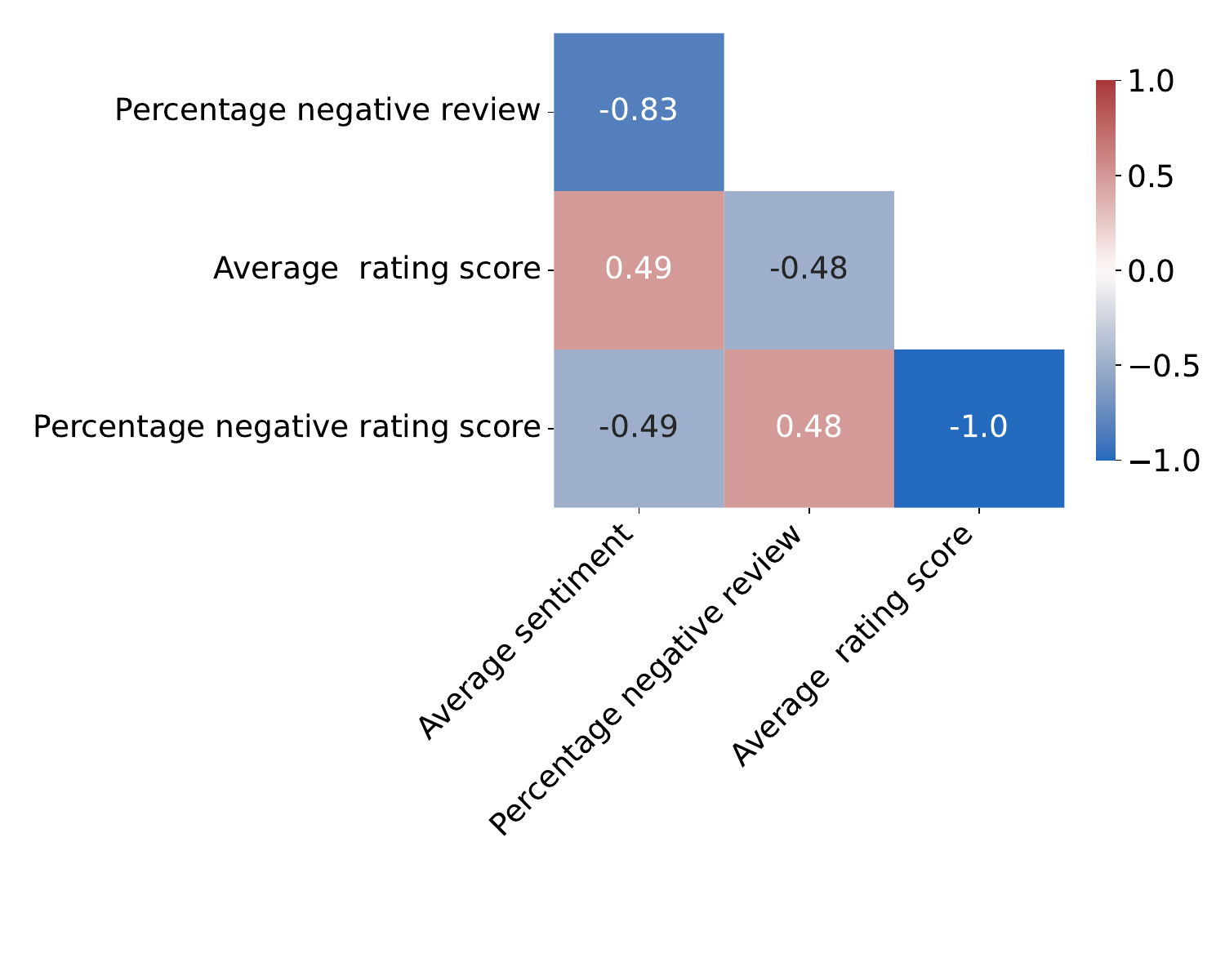}
        \caption{Before outlier removal}
        \label{fig:cor-before}
    \end{subfigure}
    \hfill
    \begin{subfigure}[t]{0.75\textwidth}
        \centering
        \includegraphics[width=\textwidth]{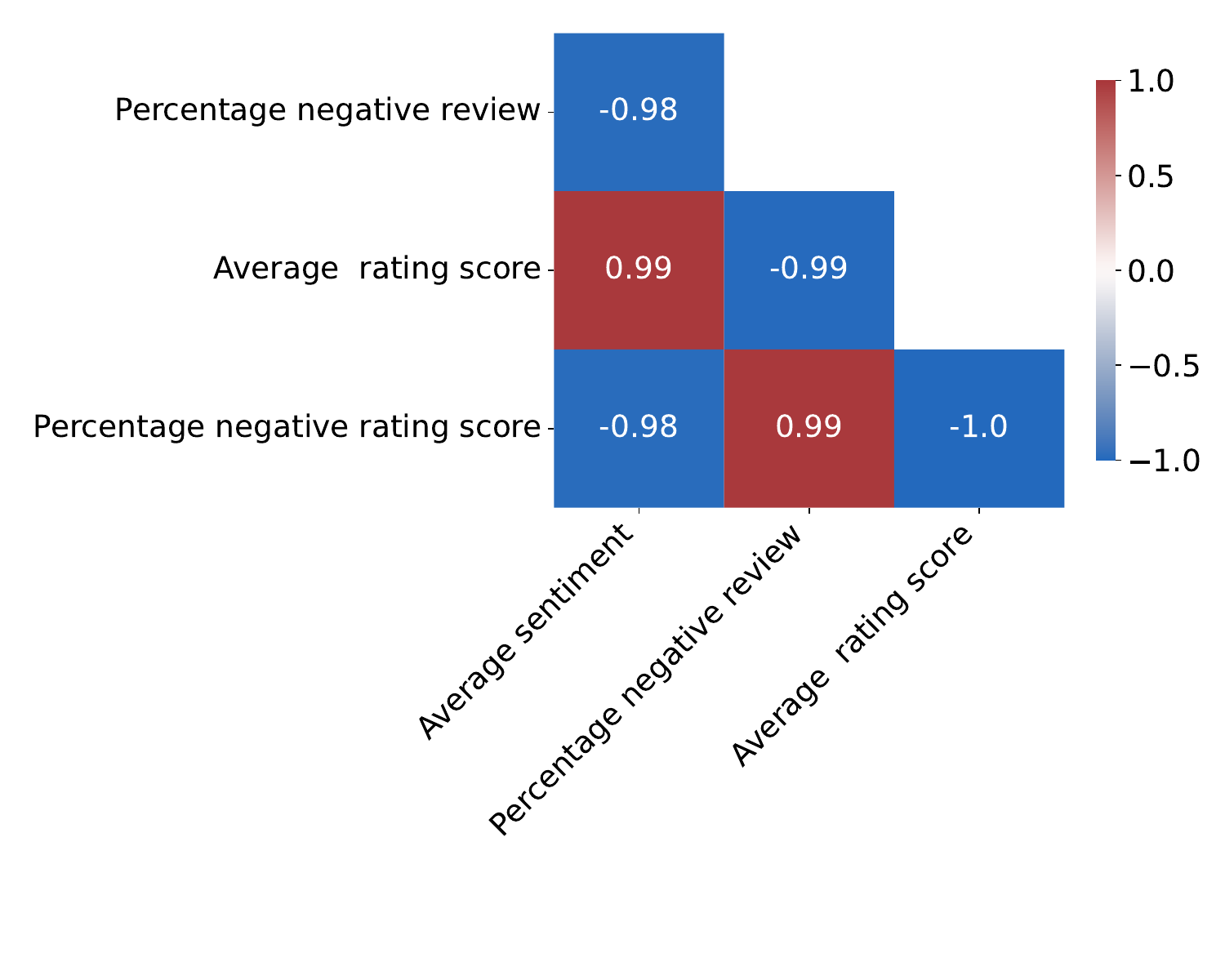}
        \caption{After outlier removal}
        \label{fig:cor-after}
    \end{subfigure}
    
    \caption{Pairwise Spearman correlation between Average sentiment, Percentage of negative reviews, Average rating score and Percentage of negative rating score of the reviews of each apps}
    \label{fig:mainfig}
\end{figure}

\textbf{Reviews with negative rating scores are highly correlated with negative sentiment reviews.}
In Fig \ref{fig:mainfig}, we analyzed Spearman's rank correlation to understand the correlation between the metrics such as \textit{Average user rating}, \textit{Average sentiment}, \textit{Percentage negative rating}, \textit{Percentage negative sentiment} before and after the outlier removal.
The results show that user ratings are highly correlated with their sentiment in the corresponding reviews. For example, the percentage of negative sentiment reviews and the percentage of negative rating scores of the studied apps have a correlation of 0.99 (p $<$ 0.001). Besides, the average sentiment of the reviews and the average rating scores of the studied apps have a correlation of 0.99 (p $<$ 0.001).
% We are trying to analyze the sentiment of the reviews to make more sense of the data. Sentiment analysis can be an add-on in interpreting  the sense of what users are complain about.
%\textbf{While most of the reviews have a positive sentiment, a negative sentiment is much more likely than a positive sentiment to be associated with a negative rating.}
% a) \textbf{User ratings and sentiments}:  We analyzed the relationship between the sentiment and the user rating provided by the users. 
 %\heng{according to here, the figure results seem to be before removing the outliers}. \bhagya{should I add the percentages after removing the outliers too?? or this is enough?}

%Therefore, we use sentiment to discuss users' complained aspects (topics) below.

%---------------------------

% \begin{figure}[!t]%{0.49\textwidth}
%          \centering
%          \includegraphics[width=.85\textwidth]{imgs/correlation_before_removal.pdf}
%          \caption{Pairwise Spearman correlation between Average sentiment, Percentage of negative reviews, Average rating score and Percentage of negative rating score of the reviews of each apps(before removing outliers).}
%          \label{correlation+}
%      \end{figure}
% \begin{figure}[!t]%{0.49\textwidth}
%          \centering
%          \includegraphics[width=.85\textwidth]{imgs/correlation_new_size2.pdf}
%          \caption{Pairwise Spearman correlation between Average sentiment, Percentage of negative reviews, Average rating score and Percentage of negative rating score of the reviews of each apps(after removing outliers).}
%          \label{correlation+}
%      \end{figure}

%\textbf{\textit{Significant user sentiment is positive for reviews.}}Most of the reviews that we are analysing and the sentiment associated with them are highly correlated and distributed at the right end of the graphs and most of the review sentiment are of positive which confirms the user rating distribution in Fig \ref{rating-distri}. \bhagya{not sure to keep here or in the discussion} All the previous studies have equal distribution of data for both positive and negative sentiment reviews, however in our condition the data is too imbalanced and data is skewed towards the positive sentiments. Which makes our data for complain factors narrower. However understanding the minority complaints are far important for the stakeholders and developers to understand the application.

\textbf{Users are mostly complaining about the advertisements and the quality (bugs, content quality, or streaming quality) of sports apps.}
%Except \emph{advertisement}, \emph{bug reports}, and \emph{negative feedback}  which are expected to have negative reviews, the review topics \emph{quality of content}, \emph{version update}, \emph{return on investment}, and \emph{streaming quality} are also among the topics mostly associated with negative reviews.
Table~\ref{tab:allsentireview} shows the average sentiment of the reviews associated with each topic and the percentage of reviews with negative sentiment (i.e., score $<=$ -0.05), as well as the average ratings and the percentage of reviews with a negative rating (rating $<=$ 2) associated with each topic.
% , most of which are not identified in prior studies that analyze the user-complaints of general mobile apps~\cite{fu2013people,mcilroy2016analyzing}.
As expected, the reviews associated with the topics of \emph{advertisement}, \emph{bug reports}, and \emph{negative feedback} have the lowest average sentiment, which confirms the results of prior studies~\cite{kalaichelavan2020people,fu2013people,mcilroy2016analyzing}. Among the other review topics,  the ones with the lowest average sentiment values include the topics such as
\emph{quality of content}, \emph{version update}, \emph{return on investment}, and \emph{streaming quality}.
The topics of \emph{quality of content} and \emph{version updates} have been identified in prior studies as topics~\cite{kalaichelavan2020people,fu2013people,mcilroy2016analyzing}; however, they are not identified as negative aspects in these studies. %for the users of the given study. 
Our results indicate that these are specific to sports apps. The topics \emph{return on investment} and \emph{streaming quality} are also recognized for generic apps as negative sentiment topics \cite{fu2013people}. 

The following are the review topics associated with the most negative sentiment:

\textit{Advertisement}: 76\% of the apps that we analyzed are free applications (completely or with premium purchases inside). Most of the revenue of these applications comes from advertisements.
Our results show the negative implications of having advertisements on user experiences. Thus, the question of compromising user satisfaction for income needs to be considered by developers. 

\textit{Bug Reports}: Irrespective of the challenges with natural language analysis, some users are very intricate about their issues and inform it via reviews. The sentiment associated with the reviews indicates that they are not happy about it(\ref{tab:topic-summary}) however, users provide detailed issues, including how to replicate and what is the expected behaviour-- which helps the developers mitigate the issue faster.

\textit{Quality of Content}: Table \ref{tab:topic-summary} shows that this is the second most popular topic among the reviews. %, which spread across different functionalities\heng{Table 4 does not show anything related to functionalities?}. 
From manually examining the reviews, it is clear that \textbf{both subjective (e.g., content appropriate to user groups) and objective (e.g., network metrics and HTTPS streaming metrics)} aspects are user concerns that developers have to put more focus on. 

\textit{Negative Feedback:} This is a prominent topic with an expected negative sentiment, which demonstrates the validity of our topic modeling results~\cite{schobel2013using}.
%proves our topic modeling is more accurate and our classification is valid~\cite{schobel2013using}.
%\heng{Check if these topics are identified as negative topics/aspects in prior works: if yes, add citations; of no, say that it's new finding for the specific sports app category.}\bhagya{updated}
%can find related topics in prior studies (\emph{additional cost} in \cite{mcilroy2016analyzing} and \emph{media},\emph{version} in \cite{fu2013people}, respectively).

\begin{table}[]
\caption{Average sentiment and average rating score of reviews with distribution of negative sentiment reviews and negative rated reviews across different topics. Table is sorted based on the the distribution of negative sentiment review distribution across topics 
% \heng{put the table in results} \heng{it looks strange that Advertisement has one of the best sentiment, and positive feedback has a high percentage of negative reviews, please double check the result}}\bhagya{I have looked into the results, there are reviews with neutral and positive feedback with a mention of the issues with the advertisement}\bhagya{similarly the issue with the positive feedback is the category of the apps with highest number of reviews- thus the the number of negative sentiment is also higher ( issues with imbalanced class)} \heng{The number of the reviews does not matter, as you are measuring the percentage, right? Comparing this table 4 and table 3 in RQ2, it seems the percentage of negative reviews here is highly correlated with the topic percentage in Table 3. Are you calculating the percentage here this way: for example, for the ``positive fedback'' topic, divide the number of negative reviews with the ``positive feedback'' by the number of reviews with the ``positive feedback'' topic? It seems something is wrong here. }
}
\label{tab:allsentireview}
\centering
\scalebox{0.6}{
\begin{tabular}{@{}|l|c|c|c|c|@{}}
\toprule
\textbf{Topics} & \multicolumn{1}{l|}{\textbf{\begin{tabular}[c]{@{}l@{}}Percentage negative \\ sentiment review\end{tabular}}} & \multicolumn{1}{l|}{\textbf{Average sentiment}} & \multicolumn{1}{l|}{\textbf{\begin{tabular}[c]{@{}l@{}}Percentage negative \\ rating score\end{tabular}}} & \multicolumn{1}{l|}{\textbf{Average  rating score}} \\ \midrule
\textbf{Advertisement} & 35.66\% & -0.02 & 59.60\% & 2.4 \\ \midrule
\textbf{Bug reports} & 33.67\% & 0.06 & 50.90\% & 2.71 \\ \midrule
\textbf{Quality of content} & 30.48\% & 0.17 & 45.40\% & 2.94 \\ \midrule
\textbf{Negative feedback} & 29.02\% & 0.16 & 35.86\% & 3.31 \\ \midrule
\textbf{Version update} & 21.72\% & 0.23 & 26.33\% & 3.54 \\ \midrule
\textbf{Return on investment} & 20.28\% & 0.31 & 27.87\% & 3.65 \\ \midrule
\textbf{Streaming quality} & 19.78\% & 0.3 & 25.62\% & 3.78 \\ \midrule
\textbf{Feature request} & 10.74\% & 0.42 & 14.81\% & 4.18 \\ \midrule
\textbf{Precision of tools} & 7.35\% & 0.52 & 8.10\% & 4.37 \\ \midrule
\textbf{Accuracy of prediction} & 7.27\% & 0.43 & 9.98\% & 4.29 \\ \midrule
\textbf{Impact of application} & 4.42\% & 0.56 & 3.23\% & 4.63 \\ \midrule
\textbf{Up-to-dateness} & 3.33\% & 0.52 & 4.32\% & 4.63 \\ \midrule
\textbf{Review credibility} & 2.51\% & 0.6 & 3.68\% & 4.71 \\ \midrule
\textbf{Positive feedback} & 2.51\% & 0.51 & 4.08\% & 4.68 \\ \bottomrule
\end{tabular}}

\end{table}

% \input{texfiles/images/sentiment-function-topic}
% \input{texfiles/images/sentiment-func-topic-updated}
% \usepackage{booktabs}

%\begin{table*}[!t]
\begin{sidewaystable}
%\begin{adjustbox}{angle=90} 
\centering
\small

\caption{Distribution of average review sentiment over different topics and different functionalities of sports apps. Also the percentage of negative sentiment reviews out of all reviews.} %\heng{all sentiment values are positive. the results are not quite helpful for answering the RQ: what are people complaining about? In addition to the average sentiment values, add the percentage of reviews with negative sentiment (what's the threshold used for tagging a negative sentiment, 0.05?). For example, put 10.23\% (0.20) where 10.23\% indicates the percentage of reviews of a app category and a topic that have negative sentiment, 0.20 indicate the average sentiment of all reviews of the app category and the topic.}
\label{tab:topic-function-senti}
%\heng{for each app category, highlight the top three topics with highest percentage of reviews with negative sentiment}
%\heng{how to explain the positive feedback topic has the highest percentage of negative reviews? double check the result}
%\heng{here the grand total are for the sampled apps right? need a separate table for the topic sentiment of all 2,058 apps, similar to this table but only showing overall topic - sentiment (average sentiment and percentage of negative reviews). Then the grand total row can be removed from here (duplicated information).}\bhagya{I don't think it will be a duplicate information rather it shows the difference in distribution with the sample set and the whole dataset}\bhagya{updated rest of the comments}
%\heng{similar comments as for Table 4: check the results }
%\bhagya{updated}
\scalebox{0.55}{
\begin{tabular}{@{}lccccccccccccccc@{}}
\cmidrule(r){1-14} \cmidrule(l){16-16}
\multicolumn{1}{|l|}{\textbf{Topic name}} & \multicolumn{1}{l|}{\textbf{Betting}} & \multicolumn{1}{l|}{\textbf{Betting Tips}} & \multicolumn{1}{l|}{\textbf{Health Tips}} & \multicolumn{1}{l|}{\textbf{\begin{tabular}[c]{@{}l@{}}League \\ Management\end{tabular}}} & \multicolumn{1}{l|}{\textbf{Live updates}} & \multicolumn{1}{l|}{\textbf{News}} & \multicolumn{1}{l|}{\textbf{Radio}} & \multicolumn{1}{l|}{\textbf{\begin{tabular}[c]{@{}l@{}}Social \\ Network\end{tabular}}} & \multicolumn{1}{l|}{\textbf{Streaming}} & \multicolumn{1}{l|}{\textbf{\begin{tabular}[c]{@{}l@{}}Team \\ management\end{tabular}}} & \multicolumn{1}{l|}{\textbf{Tools}} & \multicolumn{1}{l|}{\textbf{Tracking}} & \multicolumn{1}{l|}{\textbf{Training}} & \multicolumn{1}{l|}{\textbf{}} & \multicolumn{1}{l|}{\textbf{Grand Total}} \\ \cmidrule(r){1-14} \cmidrule(l){16-16} 
\multicolumn{1}{|l|}{\textbf{\begin{tabular}[c]{@{}l@{}}Accuracy \\ of prediction\end{tabular}}} & \multicolumn{1}{c|}{0.38(13.1\%)} & \multicolumn{1}{c|}{0.49(5.71\%)} & \multicolumn{1}{c|}{0.46(2.53\%)} & \multicolumn{1}{c|}{0.54(0\%)} & \multicolumn{1}{c|}{0.38(4.48\%)} & \multicolumn{1}{c|}{0.43(6.12\%)} & \multicolumn{1}{c|}{0.21(12.12\%)} & \multicolumn{1}{c|}{0.4(14.29\%)} & \multicolumn{1}{c|}{0.06(22.39\%)} & \multicolumn{1}{c|}{0.29(20\%)} & \multicolumn{1}{c|}{0.33(4.55\%)} & \multicolumn{1}{c|}{0.47(7.41\%)} & \multicolumn{1}{c|}{0.41(9.64\%)} & \multicolumn{1}{c|}{} & \multicolumn{1}{c|}{\textbf{0.43(7.27\%)}} \\ \cmidrule(r){1-14} \cmidrule(l){16-16} 
\multicolumn{1}{|l|}{\textbf{\begin{tabular}[c]{@{}l@{}}Advertise-\\ ment\end{tabular}}} & \multicolumn{1}{c|}{\textbf{-0.06(33.33\%)}} & \multicolumn{1}{c|}{\textbf{0.08(25.33\%)}} & \multicolumn{1}{c|}{\textbf{0.09(30.77\%)}} & \multicolumn{1}{c|}{-0.07(40\%)} & \multicolumn{1}{c|}{\textbf{0.06(24.91\%)}} & \multicolumn{1}{c|}{\textbf{0.08(27.05\%)}} & \multicolumn{1}{c|}{\textbf{-0.06(51.16\%)}} & \multicolumn{1}{c|}{\textbf{-0.17(53.33\%)}} & \multicolumn{1}{c|}{\textbf{-0.11(45.66\%)}} & \multicolumn{1}{c|}{\textbf{-0.12(46.51\%)}} & \multicolumn{1}{c|}{0.04(16.67\%)} & \multicolumn{1}{c|}{\textbf{0.17(21.13\%)}} & \multicolumn{1}{c|}{\textbf{0.25(20.86\%)}} & \multicolumn{1}{c|}{} & \multicolumn{1}{c|}{\textbf{-0.02(35.66\%)}} \\ \cmidrule(r){1-14} \cmidrule(l){16-16} 
\multicolumn{1}{|l|}{\textbf{\begin{tabular}[c]{@{}l@{}}Bug \\ reports\end{tabular}}} & \multicolumn{1}{c|}{\textbf{-0.03(36.84\%)}} & \multicolumn{1}{c|}{\textbf{0.18(20.3\%)}} & \multicolumn{1}{c|}{0.39(0\%)} & \multicolumn{1}{c|}{0.07(33.33\%)} & \multicolumn{1}{c|}{0.24(14.19\%)} & \multicolumn{1}{c|}{\textbf{0.18(26.02\%)}} & \multicolumn{1}{c|}{0.24(21.67\%)} & \multicolumn{1}{c|}{0.39(8.33\%)} & \multicolumn{1}{c|}{\textbf{-0.09(47.49\%)}} & \multicolumn{1}{c|}{\textbf{-0.03(37.54\%)}} & \multicolumn{1}{c|}{0.13(28.57\%)} & \multicolumn{1}{c|}{\textbf{0.25(21.97\%)}} & \multicolumn{1}{c|}{\textbf{0.22(22.16\%)}} & \multicolumn{1}{c|}{} & \multicolumn{1}{c|}{\textbf{0.06(33.67\%)}} \\ \cmidrule(r){1-14} \cmidrule(l){16-16} 
\multicolumn{1}{|l|}{\textbf{\begin{tabular}[c]{@{}l@{}}Feature \\ Request\end{tabular}}} & \multicolumn{1}{c|}{0.3(19.51\%)} & \multicolumn{1}{c|}{0.43(3.8\%)} & \multicolumn{1}{c|}{0.49(1.61\%)} & \multicolumn{1}{c|}{0.34(14.29\%)} & \multicolumn{1}{c|}{0.42(4.17\%)} & \multicolumn{1}{c|}{0.47(8.79\%)} & \multicolumn{1}{c|}{0.54(6.38\%)} & \multicolumn{1}{c|}{0.62(8.57\%)} & \multicolumn{1}{c|}{0.22(17.48\%)} & \multicolumn{1}{c|}{0.43(12.11\%)} & \multicolumn{1}{c|}{0.44(5.94\%)} & \multicolumn{1}{c|}{0.61(3.03\%)} & \multicolumn{1}{c|}{0.55(3.85\%)} & \multicolumn{1}{c|}{} & \multicolumn{1}{c|}{\textbf{0.42(10.74\%)}} \\ \cmidrule(r){1-14} \cmidrule(l){16-16} 
\multicolumn{1}{|l|}{\textbf{\begin{tabular}[c]{@{}l@{}}Impact of \\ Application\end{tabular}}} & \multicolumn{1}{c|}{0.31(9.52\%)} & \multicolumn{1}{c|}{0.37(6.35\%)} & \multicolumn{1}{c|}{0.61(0.81\%)} & \multicolumn{1}{c|}{0.23(0\%)} & \multicolumn{1}{c|}{0.35(5.26\%)} & \multicolumn{1}{c|}{0.44(4.29\%)} & \multicolumn{1}{c|}{0.43(16.67\%)} & \multicolumn{1}{c|}{No reviews} & \multicolumn{1}{c|}{0(28.36\%)} & \multicolumn{1}{c|}{0.4(10\%)} & \multicolumn{1}{c|}{0.27(20\%)} & \multicolumn{1}{c|}{0.59(4\%)} & \multicolumn{1}{c|}{0.58(3.95\%)} & \multicolumn{1}{c|}{} & \multicolumn{1}{c|}{\textbf{0.56(4.42\%)}} \\ \cmidrule(r){1-14} \cmidrule(l){16-16} 
\multicolumn{1}{|l|}{\textbf{\begin{tabular}[c]{@{}l@{}}Negative \\ Feedback\end{tabular}}} & \multicolumn{1}{c|}{-0.12(50\%)} & \multicolumn{1}{c|}{\textbf{0.24(20.08\%)}} & \multicolumn{1}{c|}{\textbf{0.45(8\%)}} & \multicolumn{1}{c|}{\textbf{-0.28(54.55\%)}} & \multicolumn{1}{c|}{\textbf{0.16(20.63\%)}} & \multicolumn{1}{c|}{0.28(24.14\%)} & \multicolumn{1}{c|}{\textbf{0.05(48.15\%)}} & \multicolumn{1}{c|}{\textbf{0.02(33.33\%)}} & \multicolumn{1}{c|}{\textbf{-0.23(57.68\%)}} & \multicolumn{1}{c|}{\textbf{-0.13(55.41\%)}} & \multicolumn{1}{c|}{\textbf{0(37.84\%)}} & \multicolumn{1}{c|}{0.35(16.48\%)} & \multicolumn{1}{c|}{0.44(11.64\%)} & \multicolumn{1}{c|}{} & \multicolumn{1}{c|}{\textbf{0.16(29.02\%)}} \\ \cmidrule(r){1-14} \cmidrule(l){16-16} 
\multicolumn{1}{|l|}{\textbf{\begin{tabular}[c]{@{}l@{}}Positive \\ Feedback\end{tabular}}} & \multicolumn{1}{c|}{0.53(2.44\%)} & \multicolumn{1}{c|}{0.53(0.7\%)} & \multicolumn{1}{c|}{0.51(0.2\%)} & \multicolumn{1}{c|}{0.34(11.76\%)} & \multicolumn{1}{c|}{0.52(0.62\%)} & \multicolumn{1}{c|}{0.6(1.98\%)} & \multicolumn{1}{c|}{0.57(1.33\%)} & \multicolumn{1}{c|}{0.52(4.55\%)} & \multicolumn{1}{c|}{0.22(17.63\%)} & \multicolumn{1}{c|}{0.53(1.25\%)} & \multicolumn{1}{c|}{0.5(1.67\%)} & \multicolumn{1}{c|}{0.61(1.34\%)} & \multicolumn{1}{c|}{0.55(0.57\%)} & \multicolumn{1}{c|}{} & \multicolumn{1}{c|}{\textbf{0.51(2.51\%)}} \\ \cmidrule(r){1-14} \cmidrule(l){16-16} 
\multicolumn{1}{|l|}{\textbf{\begin{tabular}[c]{@{}l@{}}Precision \\ of Tools\end{tabular}}} & \multicolumn{1}{c|}{0.27(19.51\%)} & \multicolumn{1}{c|}{0.35(3.03\%)} & \multicolumn{1}{c|}{\textbf{0.45(11.11\%)}} & \multicolumn{1}{c|}{0.49(0\%)} & \multicolumn{1}{c|}{0.36(6.29\%)} & \multicolumn{1}{c|}{0.42(9.75\%)} & \multicolumn{1}{c|}{0.42(15.79\%)} & \multicolumn{1}{c|}{0.47(0\%)} & \multicolumn{1}{c|}{0.16(25.3\%)} & \multicolumn{1}{c|}{0.45(8.7\%)} & \multicolumn{1}{c|}{0.43(10.7\%)} & \multicolumn{1}{c|}{0.65(4.83\%)} & \multicolumn{1}{c|}{0.47(5.22\%)} & \multicolumn{1}{c|}{} & \multicolumn{1}{c|}{\textbf{0.52(7.35\%)}} \\ \cmidrule(r){1-14} \cmidrule(l){16-16} 
\multicolumn{1}{|l|}{\textbf{\begin{tabular}[c]{@{}l@{}}Quality \\ of Content\end{tabular}}} & \multicolumn{1}{c|}{\textbf{-0.02(27.78\%)}} & \multicolumn{1}{c|}{0.21(4.35\%)} & \multicolumn{1}{c|}{0.22(0\%)} & \multicolumn{1}{c|}{\textbf{-0.05(100\%)}} & \multicolumn{1}{c|}{0.28(6.78\%)} & \multicolumn{1}{c|}{0.31(13.79\%)} & \multicolumn{1}{c|}{0.2(25\%)} & \multicolumn{1}{c|}{-0.37(50\%)} & \multicolumn{1}{c|}{0.16(32.71\%)} & \multicolumn{1}{c|}{-0.06(29.41\%)} & \multicolumn{1}{c|}{\textbf{0.05(37.5\%)}} & \multicolumn{1}{c|}{\textbf{0.35(17.39\%)}} & \multicolumn{1}{c|}{\textbf{0.19(17.07\%)}} & \multicolumn{1}{c|}{} & \multicolumn{1}{c|}{\textbf{0.17(30.48\%)}} \\ \cmidrule(r){1-14} \cmidrule(l){16-16} 
\multicolumn{1}{|l|}{\textbf{\begin{tabular}[c]{@{}l@{}}Return on \\ Investment\end{tabular}}} & \multicolumn{1}{c|}{0.18(22.81\%)} & \multicolumn{1}{c|}{0.32(18.89\%)} & \multicolumn{1}{c|}{0.41(6.06\%)} & \multicolumn{1}{c|}{0.52(18.18\%)} & \multicolumn{1}{c|}{\textbf{0.22(15.29\%)}} & \multicolumn{1}{c|}{0.37(14.04\%)} & \multicolumn{1}{c|}{\textbf{0.32(25.45\%)}} & \multicolumn{1}{c|}{0.38(28.57\%)} & \multicolumn{1}{c|}{0.1(32.83\%)} & \multicolumn{1}{c|}{0.11(35.82\%)} & \multicolumn{1}{c|}{0.28(28.26\%)} & \multicolumn{1}{c|}{0.48(14.01\%)} & \multicolumn{1}{c|}{0.45(11.85\%)} & \multicolumn{1}{c|}{} & \multicolumn{1}{c|}{\textbf{0.31(20.28\%)}} \\ \cmidrule(r){1-14} \cmidrule(l){16-16} 
\multicolumn{1}{|l|}{\textbf{\begin{tabular}[c]{@{}l@{}}Review \\ credibility\end{tabular}}} & \multicolumn{1}{c|}{0.62(1.96\%)} & \multicolumn{1}{c|}{0.6(1.93\%)} & \multicolumn{1}{c|}{0.65(1.89\%)} & \multicolumn{1}{c|}{0.61(5.56\%)} & \multicolumn{1}{c|}{0.56(2.33\%)} & \multicolumn{1}{c|}{0.64(1.73\%)} & \multicolumn{1}{c|}{0.66(1.64\%)} & \multicolumn{1}{c|}{0.63(0\%)} & \multicolumn{1}{c|}{0.44(9.41\%)} & \multicolumn{1}{c|}{0.64(0.96\%)} & \multicolumn{1}{c|}{0.59(4.26\%)} & \multicolumn{1}{c|}{0.6(2.09\%)} & \multicolumn{1}{c|}{0.65(0.79\%)} & \multicolumn{1}{c|}{} & \multicolumn{1}{c|}{\textbf{0.6(2.51\%)}} \\ \cmidrule(r){1-14} \cmidrule(l){16-16} 
\multicolumn{1}{|l|}{\textbf{\begin{tabular}[c]{@{}l@{}}Up-to-\\dateness\end{tabular}}} & \multicolumn{1}{c|}{0.28(16.88\%)} & \multicolumn{1}{c|}{0.43(2.22\%)} & \multicolumn{1}{c|}{0.57(0\%)} & \multicolumn{1}{c|}{0.44(0\%)} & \multicolumn{1}{c|}{0.3(8.15\%)} & \multicolumn{1}{c|}{0.43(12.76\%)} & \multicolumn{1}{c|}{0.37(17.89\%)} & \multicolumn{1}{c|}{0.48(25\%)} & \multicolumn{1}{c|}{0.11(33.49\%)} & \multicolumn{1}{c|}{0.36(14.77\%)} & \multicolumn{1}{c|}{\textbf{0.29(29.41\%)}} & \multicolumn{1}{c|}{0.48(3.17\%)} & \multicolumn{1}{c|}{0.33(13.43\%)} & \multicolumn{1}{c|}{} & \multicolumn{1}{c|}{\textbf{0.3(19.78\%)}} \\ \cmidrule(r){1-14} \cmidrule(l){16-16} 
\multicolumn{1}{|l|}{\textbf{\begin{tabular}[c]{@{}l@{}}Version \\ update\end{tabular}}} & \multicolumn{1}{c|}{0.21(10.53\%)} & \multicolumn{1}{c|}{0.41(0.99\%)} & \multicolumn{1}{c|}{0.4(0\%)} & \multicolumn{1}{c|}{\textbf{-0.08(50\%)}} & \multicolumn{1}{c|}{0.46(2.58\%)} & \multicolumn{1}{c|}{0.54(3.19\%)} & \multicolumn{1}{c|}{0.66(0\%)} & \multicolumn{1}{c|}{0.35(0\%)} & \multicolumn{1}{c|}{0.19(11.21\%)} & \multicolumn{1}{c|}{0.37(13.33\%)} & \multicolumn{1}{c|}{0.29(8.33\%)} & \multicolumn{1}{c|}{0.41(9.52\%)} & \multicolumn{1}{c|}{0.43(0\%)} & \multicolumn{1}{c|}{} & \multicolumn{1}{c|}{\textbf{0.52(3.33\%)}} \\ \cmidrule(r){1-14} \cmidrule(l){16-16} 
\multicolumn{1}{|l|}{\textbf{\begin{tabular}[c]{@{}l@{}}Streaming \\ quality\end{tabular}}} & \multicolumn{1}{c|}{0.13(25.18\%)} & \multicolumn{1}{c|}{0.26(9.29\%)} & \multicolumn{1}{c|}{0.52(0\%)} & \multicolumn{1}{c|}{0.44(12.5\%)} & \multicolumn{1}{c|}{0.3(10.42\%)} & \multicolumn{1}{c|}{\textbf{0.22(26.66\%)}} & \multicolumn{1}{c|}{0.27(11.54\%)} & \multicolumn{1}{c|}{0.28(21.74\%)} & \multicolumn{1}{c|}{0.01(37.04\%)} & \multicolumn{1}{c|}{0.13(32.41\%)} & \multicolumn{1}{c|}{0.18(17.14\%)} & \multicolumn{1}{c|}{0.36(15.44\%)} & \multicolumn{1}{c|}{0.36(10.26\%)} & \multicolumn{1}{c|}{} & \multicolumn{1}{c|}{\textbf{0.23(21.72\%)}} \\ \cmidrule(r){1-14} \cmidrule(l){16-16} 
\textbf{} &  &  &  &  &  &  &  &  &  &  &  &  &  & \multicolumn{1}{l}{} & \multicolumn{1}{l}{\textbf{}} \\ \cmidrule(r){1-14} \cmidrule(l){16-16} 
\multicolumn{1}{|c|}{\textbf{Grand Total}} & \multicolumn{1}{c|}{\textbf{0.23(21.23\%)}} & \multicolumn{1}{c|}{\textbf{0.46(5.36\%)}} & \multicolumn{1}{c|}{\textbf{0.52(1.75\%)}} & \multicolumn{1}{c|}{\textbf{0.32(17.65\%)}} & \multicolumn{1}{c|}{\textbf{0.43(5.05\%)}} & \multicolumn{1}{c|}{\textbf{0.47(8.25\%)}} & \multicolumn{1}{c|}{\textbf{0.38(16.71\%)}} & \multicolumn{1}{c|}{\textbf{0.4(15.15\%)}} & \multicolumn{1}{c|}{\textbf{0.1(32.12\%)}} & \multicolumn{1}{c|}{\textbf{0.33(18.18\%)}} & \multicolumn{1}{c|}{\textbf{0.34(15.14\%)}} & \multicolumn{1}{c|}{\textbf{0.56(7.18\%)}} & \multicolumn{1}{c|}{\textbf{0.54(4.52\%)}} & \multicolumn{1}{c|}{\textbf{}} & \multicolumn{1}{c|}{\textbf{}} \\ \cmidrule(r){1-14} \cmidrule(l){16-16} 
\end{tabular}
}
\label{tab:sent-function-topic}
\end{sidewaystable}

\textbf{\textit{Streaming}, \textit{Betting}, \textit{Team management} and \textit{League management} are the most complained sports app functionalities.} 
Table~\ref{tab:sent-function-topic} shows that the percentage of negative sentiment reviews is 32.12\%, 21.23\%, 18.18\% and 17.65\% for the apps associated with the functionalities of \textit{streaming}, \textit{Betting}, \textit{Team management} and \textit{League management}, respectively. 
The most complained topics of these apps include \textit{advertisement}, \textit{bug reports}, \textit{quality of content}, and \textit{negative feedback}, and \textit{version update}. 
\review{1.2}{The findings suggest that users prioritize their concerns regarding the advertisement and quality of functionalities supported by sports apps, including bugs or the quality of content. Our analysis highlights the most frustrating topics among competitors, which developers can use to prioritize their efforts in addressing these issues and ensure user retention. }

\textbf{In general, the most complained topics are consistent across the sports apps with different functionalities.}
Table~\ref{tab:sent-function-topic} shows the review sentiment for each review topic of each app functionality. 
The topics \text{advertisement}, \textit{bug reports}, \textit{negative feedback}, and \textit{quality of content} are consistently among the most complained topics of each functionality. 
For example, 
these four topics are the most complained topics of the most popular functionalities (Table~\ref{rq1results}): Betting Tips and Training apps.
%Live updates and Streaming. We identified the most appeared functionalities; for example, \textit{Betting Tips} the most negative sentiment topics are \textit{Advertisement}(25.33\%), \textit{Bug reports}(20.3\%) and \textit{Negative feedback}(20.08\%). Similarly, for \textit{Streaming} \textit{Advertisement}(45.66\%), \textit{Bug reports}(47.49\%) and \textit{Negative feedback}(57.68\%) but has a higher percentage of negative reviews with average sentiment less than -0.05. \textit{Training} top two negative sentiment are the same as other functionalities \textit{Advertisement}(20.86\%), \textit{Bug reports}(22.16\%) followed by \textit{Quality of content}((17.07\%). 
However, there are some exceptions. For example, the topic of \textit{streaming quality} is among the most complained topics of the \textit{News} apps, which indicates users' concerns regarding the quality of \textit{News} apps' multi-media streams. %while \textit{return on investment} is among the most complained topics of the \textit{Live updates} apps.
%However, for the \textit{Live updates} along with \textit{Advertisement}(24.92\%) and \textit{Negative feedback}(20.63\%) \textit{Return on investment}(15.29\%).

% \textbf{Our results indicate analyzing the reviews of a large number of apps in a specific app category can provide new insights about what users complain about an app category}. 

% \heng{the text below is repeating previous content or no?}
% We continued our analysis on a more qualitative level, where we looked into the main functionality of the apps and how the average sentiment and percentage of negative sentiment values obtained are associated with them. The most important functionalities such as Streaming, \textit{Team management} and \textit{League management} have paid purchases on premium packages. Streaming errors are mostly located on the apps with main functionality as video streaming, radio broadcasting and news. And the complaints are about the quality of streaming on HTTP services \cite{rodriguez2014impact}. Negative feedback and Bug reports are distributed over all the main functionalities. The most available streaming apps are for the sports with the most fans, football and cricket. 
% As shown in the Figure ~\ref{correlation+}, the average rating score and the average sentiment positively correlated with a coefficient of 0.99. In contrast, the percentage of negative rating score and the percentage of negative reviews are negatively correlated. Thus, the reviews with low ratings tend to be associated with low sentiment scores. 

\begin{tcolorbox}
%\heng{not updated and not interesting. check the bold sentences}\bhagya{updated}
Users are generally positive about sports apps, as indicated in their review sentiment. On the other hand, they are mainly complaining about the advertisements and quality (bugs, content quality, streaming quality) of these apps, which are consistent among the apps of different functionalities.
In particular, \textit{Streaming}, \textit{Betting}, \textit{Team management} and \textit{League management} are the most complained about sports app functionalities.

%General review sentiment associated with reviews are positive. Furthermore, most of the negative rated reviews are negative sentiment reviews as well. Along with \textit{Bug report} and \textit{Negative feedback}, \textit{Advertisement} and \textit{Quality of content} were most negative sentiment topics. On a functional level analysis the top functionalities \textit{Betting Tips, Training, Streaming and Live updates} contains the similar pattern of the negative rated topics. Moreover, the specific functionality studies example \textit{Team management} has \textit{Version update} as the concerned factor. It shows that developers to focus clearly on the functionalities and topics to identify user concerns and improve the user experience.
% The percentage of negative sentiment and average sentiment are closely related to the user feedback. The topics such as negative feedback and bug reports groups user concerns and the sentiment analysis correlate the user ratings and indicate the causation through the topic and functionality mapping. 
% Most concerns on the applications are on the experience of users, but the hate factors varies based on the applications are paid or free. However, users have more expectations when the applications are paid. Also users are more hateful when the user experience is lesser in those application which are time bound such as live updates.
\end{tcolorbox}

\section{Discussions} \label{sec:discussions}

%\heng{Please revisit the discussions based on the updated results and discussions. The discussions should be linked closely to the concrete findings (not the approaches) of the paper (e.g., related to the take-home messages of each RQ). Currently most of the discussions are too general and not linked to the results of the paper. Would be good to use bold sentences (central ideas) to guide each point of discussion. One and only one central point for each discussion. An example is Section 5 - Implications in paper ``A Qualitative Study of the Benefits and Costs of Logging from Developers’ Perspectives''} \bhagya{updated}

%In the previous sections, we analyzed the data extracted from Google Play Store, and the research questions were put forward in the introduction. In this section, we will evaluate the following actors; Users, Developers and Researchers in society with implications of our results. 

In this section, we discuss the implications of our results.

\review{2-1, R3-1}{\noindent\textbf{Sports apps are a rapid-growing in Google play store}. Our results in RQ1 indicate that the number of sports apps are steadily increasing in recent years. It is reported that there are over 46K sports apps on the Google Play Store \cite{AppbrainStats}. On the other hand, there are many sports apps targeting the same sports types or implementing similar functionalities. For example, only out of our sampled 336 apps, there are 40 apps and 22 apps that provide betting tips and live updates for football, respectively. The competitive nature of the sports app market suggests that developers need to pay extra attention to the user perceived quality, which is communicated through their ratings and review text, for example, paying attention to aspects complained about most by the users ( e.g, bugs and ads). From the identified topics, developers and researchers can locate specific issues that are bothering users and fix them to improve the user experience. Our results comprising of understanding user user sentiment among a functional level can provide guidance for developers to improve the user-perceived quality of their apps and win the competition in the market. For example, in a review, \emph{``Great app to keep your body fit and healthy…. However, as everyone says, the app is broken because it’s not counting completed workouts. Although it’s being synchronized with Fit correctly, progress inside the app is just not tracking itself. Please fix it. The app is great, but this bug is irritating. .”} Categorized as a bug fix, developers can identify the issues and prioritize them based on the sentiment score associated with it, which is 0.0414, indicating a neutral sentiment due to the positive first part and negative second part of the review. Our topic analysis can assign a "bug" topic, and our sentiment analysis can assign a non-positive sentiment, suggesting developers to fix the bug. We also encourage future work to study the factors that influence the competitiveness of sports apps in the market. For example, regression models can be leveraged to explain how the app features (e.g., readabilty of descriptions)  impact app ranks or ratings.  Such analyses can be performed at two levels: 1) performed at the categorical level to understand how the features impact the ranks/ratings of sports apps in general, and 2) performed at the functional level to understand how the features impact the ranks of sports apps within a mutually-competing group (e.g., football training apps).}

\noindent\review{2-1}{\textbf{To enhance the user experience, it is crucial for developers to effectively address and manage the topics or issues associated with the most negative sentiment.} Recent studies have shown that companies with Customer Experience (CX) enabled development experience better revenue growth compared to those that do not. Additionally, they provide better profits and customer experiences \cite{Genesys2021}. However, the same survey reveals that developers are not taking measures for the success of CX initiatives due to a lack of access to correlated CX outcome data \cite{Genesys2021}. Our research bridges this gap. We identified that the topics of Advertisements and Bug Reports are the most unpleasant among user reviews. Further analysis revealed that the Streaming functionality received the most reviews in these categories. To investigate the specific bugs, we narrowed our dataset to the Streaming functionality and the review topic of Bug Reports, and analyzed the reviews. The results shows that further efforts are needed to improve the quality of the apps or explore other revenue-generating options besides advertisement. The functional level analysis employed in our study proves to be useful in understanding and prioritizing issues among competitors. This same concept can be applied to improve app ratings and eventually app ranking, where app ratings and reviews are essential components of app store optimization \cite{karagkiozidou2019app}. In order to enhance the rating score of the application, developers can delve into various metadata associated with the apps. They could analyze the correlation between these metadata and the rating score at different levels such as Google Play Store, sports app category, and even on a functional level. This analysis will enable them to grasp the nuanced correlation within the diverse strata of the data. The sentiment score enabled us to prioritize which topics to address in order to improve the user experience. Our study is easy to replicate using the provided replication package and dataset that we used. Additionally, it can be applied to other domains with the minimal overhead of initial manual data labelling.}

\noindent\textbf{Sports apps provide good examples of studying the integration of data science or machine learning techniques in software applications}. As discussed in RQ1, a significant portion (37.85\%) of sports apps leverage statistical or predictive analyses (a.k.a., sports analytics). For example, sports apps providing betting tips, training, and tracking functionalities may rely on statistical or predictive analyses to help users make data-driven decisions. In RQ2, we also observe that users are concerned about the topic of the accuracy of prediction. Our findings highlight the significant applications of statistical and predictive analyses in sports apps, which suggest that sports apps may be explored in future work to study the engineering and integration of data science or machine learning techniques in real-world software applications. \review{2-1}{Specifically, the integration of data analytics in app development in a responsible manner presents a promising research avenue ~\cite{clarke2018guidelines}. One possible approach to achieve this goal is to examine the source code of the application to understand the application of machine learning practices in mobile app development. Such an analysis may also investigate the correlation between the analytical methods employed and the user experience provided by the app at a categorical (i.e., sports apps) or functional (e.g., football training apps) level.}

\noindent\review{2-1}{\textbf{Analyzing a targeted category of apps (e.g., sports apps) can provide more specific insights than analyzing apps across different categories while still being relevant for a large number of apps (e.g., tens of thousands of apps in a category.} In this work, we started a large number (2,058) of apps in sports apps in contrast to traditional large-scale studies that study apps in the app store regardless of their categories. Such a study can provide more specific insights for the app category, for example, the characteristics of the sub-categories (e.g., by sports types or by the provided functionalities), the review topics (e.g., the review topics such as accuracy of prediction, up-dateness, precision of tools that are new to the sports apps), and the aspects that users complain about. Although the results may be limited to a specific app category, as an app category typically include thousands to hundreds of thousands of apps, the results can still be relevant to a larger number of apps and developers. Functional or categorical level analysis bring more granular level data which are very specific among the competitors and helps identify specific user needs. Our analysis within the sports genre identifies specific features associated with user experience such as issues with integration with Chromebooks. Some features such as the need for a game score (e.g., live score update apps), file attachment (e.g., Team management apps, or response feature (e.g., Social Network) specific for the functionality the apps are offering. Identifying specific topics allows developers and researchers to gain insights into their applications without the burden of reading through all reviews. Furthermore, the analysis provides an understanding of the features that are lacking among competitors and highlights areas where developers can focus on to attract more customers and gain a competitive edge.
}

\noindent\review{2-1}{
\textbf{Our findings emphasize the imperative for developers to create features requested by the users in their apps apart from others in the Google Play Store.} The functional or categorical level analysis brings more granular level data which are very specific among the competitors and helps identify specific user needs. Understanding user needs and development towards improving the user experience has been recently adopted in the development lifecycle. To identify the distinctive features, we can focus on the Feature request topic that we identified. Given the example of a simple NLP technique, feature selection, where we extracted 3 grams to 7 grams and analyzed what the results associated with the keyword feature are, the developers can make sense of it for improving the application. As we mentioned in our paper, this can be further analyzed in a multilevel fashion within sports as a category (genres in the Google Play store) or within defined functionality (competitors). Our analysis within the sports genre identifies specific features associated with user experience. For example, we identified the following keywords within the topic feature requests using scikit learn\footnote{Scikit learn: \url{https://scikit-learn.org/stable/modules/feature_selection.html}} libraries feature selection functionality to identify the following, ``app compatible Chromebooks great app many features accessible'', ``response feedback feature work well design little clunky problem'', ``incisive exclusive game score feature fixture updates previews'', and ``app missing feature website like file attach''. Based on the results obtained, developers and researchers can identify compatibility features for general sports apps on Chromebooks, as well as functionality-level features such as the need for a game score, file attachment, or response feature. The identification of specific topics allows developers and researchers to gain insights into their applications without the burden of reading through all reviews. Furthermore, the analysis provides an understanding of the features that are lacking among competitors and highlights areas where developers can focus on attracting more customers and gaining a competitive edge.
} 

% Our results in RQ1 indicate that the number of sports apps are steadily increasing in recent years. 
% It is reported that there are over 46K sports apps on the Google Play Store~\cite{AppbrainStats}. On the other hand, there are many sports apps targeting the same sports types or implementing similar functionalities. %For example, only out of our sampled 336 apps, there are 40 apps and 22 apps that provide betting tips and live updates for football, respectively.%

\noindent\review{2-1}{\textbf{Translating our findings to other categories incurs minimal additional effort and resources.} The findings of our study suggest that directing attention towards a specific category can yield more detailed insights into the specific issues users face. This knowledge enables researchers and
developers to prioritize addressing the most frequently reported concerns. Additionally, our methodology and replication package could be leveraged to perform similar analyses for  other app categories, such as personalization. Although this translation may involve manual analysis for the qualitative study of the characteristics  of apps by identifying the application’s functionality and type, the second step of topic analysis can be easily translated based on the best coherent score selection method suggested in our work. The topic naming needs to be done with expertise. The sentiment analysis for identifying concerning factors from the users can be translated to any category and all the hyperparameters are mentioned in our replication package. Furthermore, the remaining steps of our approach can be integrated into any category, or even selectively applied within specific functional areas (e.g., competitors), facilitating further analysis of subcategories within the competitive landscape.}

\section{Threats to Validity} \label{sec:threats}
This section discusses threats to the validity of our study.
\subsection{Internal Validity}

%We did our code for preprocessing of data and topic modeling using python and LDA mallet library. The code might contain error which we evaluated using peer reviewing the code repositories. 
In this work, we use automated topic modeling (LDA) to derive topics from user reviews based on the co-occurrence patterns of words in the reviews. We use a coherence score to select the optimal number of topics. The automatically derived topics may not represent the exact topics in the user reviews. To mitigate the threat, we manually analyzed the representative reviews of each topic to examine the validity of the topic modeling results and merge some similar topics based on prior practices\cite{stackoverflow}.\review{2-4.5}{
The proliferation of bot-generated and paid reviews poses a significant challenge in the analysis of mobile app reviews. Google has implemented measures to address review bombarding incidents \cite{maf} and has established regulations to prevent review gating practices \cite{googlebusiness}. Nonetheless, Google’s review cleansing mechanisms might not have been effective for some of the collected reviewers that were posted earlier. In our approach, while the removal of reviews containing certain patterns may not entirely eliminate the content of paid reviews, it can help mitigate the challenges associated with bot-generated reviews to some extent.} \review{3-2}{In our RQ3, we have employed sentiment analysis utilizing the VADER model, a rule-based approach that may exhibit limitations when confronted with intricate sentence structures and the intricate nuances of human irony \cite{tymann2019gervader}. To validate our findings, we have conducted a comparative analysis with another rule-based model, namely TextBlob, but we ultimately selected VADER due to its superior performance based on our manual checking. It is important to note that VADER has demonstrated considerable efficacy when applied to shorter sentences such as reviews, VADER outperforms individual human raters (F1 Classification Accuracy = 0.96 and 0.84, respectively) \cite{hutto2014vader}. Nonetheless, future investigations can benefit from the integration of more sophisticated machine learning techniques to foster more definitive conclusions.}

\subsection{External Validity}

%The challenges in NLP, such as sentiment analysis on ironic sentences and challenges on too much topic in topic modeling, such as the selection of the number of topics, are also valid in our research. This study is only focused on the Google Play store, which already has a plethora of data to be analyzed, and the number of android users is more than IOS users. 

This work studies 2,058 sports apps which is a small subset of all the sports apps (over 46K according to AppBrain~\cite{AppbrainStats}). The scope of our study is limited by the fact Google Play limits access to the complete list of apps in an app category. To maximize the number of sports apps studied in this work, we started from the top 500 popular sports apps provided by AppBrain and used an iterative process to identify new keywords and search for additional sports apps.\review{2-5}{ The generation of sports app-related keywords for the subsequent iterative search is based on the descriptions of the top apps provided by Appbrain, thus there is a risk of bias as the descriptions of popular apps in the selected region of the study may influence the associated keywords. One of the ways to address this limitation is by replacing the app description used for keyword identification with new apps discovered from the previous search, thereby enriching the dataset. In this work, we iteratively use the descriptions of the newly identified sports apps to enrich the keywords, thus increasing the comprehensiveness of the covered sports apps.} We also manually verified each of the resulting apps to filter out the ones that do not belong to the category. Our obtained apps cover a variety of sports categories and app functionalities. 
\review{2-7}{In our RQ1, one of our objectives was to comprehend the statistical and analytical algorithms utilized in apps, as indicated by their descriptions. The aim was to explore the convergence of software engineering and data science for future studies. However, it is important to note that due to privacy policies, business interests, or limitations in word count, developers may or may not disclose this information, which can impact the outcomes of our analysis. In the future, we plan to examine the source code of the sports apps to better understand the statistical an danalytical algorithms used in these apps.}
%Our study was limited to a smaller dataset due to the time and resource limitations that we faced, but the results on the sample set are promising. Moreover, for sample set creation we made sure that we are considering enough confidence level and error values from each apps rather than the entire dataset to reduce the overloading of reviews with many review which could bias the dataset. However, we are extending this research on the entire dataset, hoping that we can replicate the results and be publicly available for further study.

\subsection{Construct Validity}

% We have manual labelling of the dataset; although we plan to introduce Flais Kappa inter-rater reliability, human biases can be present.
\review{2-9}{In our research, our choice of genre for the category is sports. The choice of the category was done by considering the following facts. Sports is in the top three categories of average daily time users spend worldwide on mobile apps from October 2020 to March 2021 \cite{statista}, and a diverse number of features and functionality exists within a genre, exhibiting varying degrees of data granularity. However, because of the diversity of functionality available in the sports category, the data granularity will not be as cohesive as a category with few functionality. But our results showed that yet it helped to derive domain-specific results. Thus future works on categories such as personalization with a limited number of functions can be compared to confirm our hypothesis. }

In RQ1, we manually studied a sample of sports apps to understand their characteristics. The results may be biased by the individuals who performed the analysis. To mitigate the bias, three researchers were involved in the analysis, and we ensured that we achieved a reliable agreement between the researchers (i.e., with a Cohen's kappa of 0.81). 
In addition, our manual labelling of the topics resulting from the topic modeling may be biased. 
For validating the topics generated from topic modeling, we consulted with a group of app developers, mainly focused on cross-platform (such as react native\footnote{https://reactnative.dev/}  - an open-source software framework by Meta for Android and iOS mobile application development and flutter\footnote{https://flutter.dev/} - an open source software development toolkit by Google for Android and iOS mobile application development). They were provided with representative words and 5 examples of each topic to validate the topic labels. They were asked to rate Y if they agreed with the label and N if they did not. The results show that they agree with our labels with 100\% positive feedback. % consistency was 100\%.

\review{3-2}{In RQ3, we study the sentiment of the reviews after removing outliers (e.g., reviews with high ratings but negative sentiments). We do so to ensure the accuracy of the sentiment values for the analyzed reviews. However, the outliers potentially encompass valuable evaluation information or inherent linguistic nuances such as irony or politeness, which remain complex facets of natural language processing.
}

\review{2-13}{\subsection{Conclusion Validity}In RQ3, our findings indicate a strong positive correlation between the percentage of negative reviews and the corresponding negative score, as well as a significant positive correlation between the average rating score and the average sentiment score. It is important to note, however, that these results were obtained after removing outliers from the dataset. The outliers potentially encompass valuable evaluation information or inherent linguistic nuances such as irony or politeness, which remain complex facets of natural language processing. Our study focused exclusively on a specific category of applications, therefore, it is important to note that our findings may not be applicable to other categories of applications. In order to validate our conclusions, further research could be conducted specifically targeting different categories of applications. In RQ3, we have examined the connection between the sentiment score attributed to each topic and the corresponding functionality of the apps. It is worth noting that the intricate nature of natural language could impact the accuracy of the topics and the sentiment metrics, hence potentially exerting cumulative effects on this relationship.
}

\section{Related work}\label{sec:related}
We discuss the related literature from two perspectives: prior works on app review analysis and prior works on sports applications.

\subsection{Prior works on review analysis on general applications}
There have been a lot of approaches to classifying (e.g., topic modeling) and understanding (e.g., sentiment analysis) user reviews. For example, 
a survey of the hate factors of the users on the application \citep{fu2013people} evaluates the reviews of play store apps across different categories. Chen et al. \citep{chen2014ar} presented topic extraction strategies for filtering reviews. Studies related to user reviews had been conducted across platforms (e.g., Apple Store, Samsung, Fdroid) \citep{vasa2012preliminary}. Another study evaluates a motive-based review analysis \citep{platzer2011opportunities} on apple store predicts the acceptance of the apps based on the reviews. Reviews as a source of ground-sourced information of applications and analyzed the issues from reviews that can help the developers\citep{gao2018online}. More studies on reviews are with different tools such as variation of LDA \citep{wei2006lda},  sentimental analysis \citep{guzman2014users,fu2013people} and summarising tools \citep{di2017surf}. There have been other approaches other than topic modeling on understanding the reviews, such as understanding the keywords \citep{vu2015mining} by understanding the vocabulary of the reviews\citep{hoon2012preliminary}. There have been other studies which use topic modeling on app-related other functions, such as analyzing the importance of peer review\citep{hassan2022importance}. 

The previous studies indicate that reviews can be useful in understanding user concerns, thus, providing insights for improving the app’s quality. %The review analysis techniques work better with the LDA topic modeling algorithm. Our study confirms and extends the review analysis along with the sentiment analysis of the reviews.
Prior studies analyze app reviews from a general perspective (across various app categories) or focus on a very small number of apps. In this paper, we show how concentrating the review analysis on a single app category (sports) which often includes a large number of apps may help to better understand the app reviews and the issues that users complain about. %the essential issues in each area. Thus we identify more specific issues related to that category (e.g., new topics) and specific concerns and user sentiment related to sports applications. %However, we mentioned works on the review analysis on the topic and the difference between the topics that have been adopted in our research and how they differ from the previous findings in Table \ref{tab:topic-summary}.

\subsection{Prior works on sports applications}

Prior works show that sports apps were a focused group for studies. One study evaluates the quality of recreational sports apps \citep{sportsapp1} by evaluating the apps using an expert panel approach. Francois et al. \citep{sportsapp2} identified the low quality of free coaching apps by considering 30 apps. The content of these apps was compared against the current guidelines and fitness principles established by the American College of Sports Medicine (ACSM). Pengcheng et al.~\citep{shen2020analysis}evaluated the need for Sports and Fitness Apps from the Perspective of “Healthy China.” The paper investigates the development status of sports and fitness apps and user satisfaction. 
%The challenges due to broad app categories made the researchers focus on a specific category for more meticulous results. 
There are further studies on the qualities of the apps. Francois et al. \citep{sportsapp4} evaluated the fitness apps’ quality with two independent raters. Another study on analyzing the potential of the fitness apps from Adria et al. \citep{sportsapp5} evaluates the validity and reliability of apps designed for CRF assessment. %They also assess the reviews and ratings of the apps.

%The previous studies indicate that the reviews can be useful in understanding user concerns, thus, improving the app’s qualities. The review analysis techniques work better with the LDA topic modeling algorithm. Our study confirms and extends the review analysis along with the sentiment analysis of the reviews. 
%Finally, sports app-related studies show that it is essential to analyze app quality. 
%The mentioned studies mainly focus on evaluating the quality with the ratings from industry experts. Our analysis also inspires the importance of assessing the quality of sports apps; we take user reviews as the ground truth than manual rating and conclude with the topic being discussed and the things the users mostly complain about.
The prior studies study sports apps and their quality from the perspectives of social standards or experts’ opinions. Besides, they often focus on a small number of sports apps. In comparison, this work performs a study that analyzes a large number (2,058) of sports apps from the perspectives of the app users (i.e., the user feedback of the apps reflected in their reviews).

\section{Conclusions} \label{sec:conclusions}

This paper studies the characteristics of a large number (2,058) of sports apps in the Google Play store and their user reviews.
Through a manual analysis, we identified sports apps that cover 16 sports types (e.g., Football, Cricket, Baseball) and 15 main functionalities (e.g., Betting, Betting Tips, Training, Tracking). Through automated topic modeling, we extracted 14 topics from the user reviews, among which 3 are specific to sports apps (\textit{accuracy of prediction}, \textit{up-to-dateness}, and \textit{precision of tools}). Finally, through sentiment analysis, we observed that users are mainly complaining about the advertisements and quality (e.g., bugs, content quality, streaming quality) of sports apps.
Through the large-scale study focusing on a single app category, we demonstrate that analyzing a targeted category of apps (e.g., sports apps) can provide more specific insights than analyzing apps across different categories while still being relevant for a large number of apps. Besides, as a rapid-growing and competitive market, sports apps provide rich opportunities for future research, for example, to study the integration of data science or machine learning techniques in software applications or to study the factors that influence the competitiveness of the apps.
We encourage future work to perform more studies on focused app categories to provide more meaningful observations and recommendations for the specific app categories.

%in the competitive market. 
%Our study identifies the characteristics of the sports apps with manual analysis followed by automatic analysis of the user reviews using topic modeling and sentiment analysis. We identified that the top sports category is football, and the most extensive functionality is betting tips and Training. Furthermore, we concluded that more than 1/4th of the apps use predictive methods. From topic modeling, we listed 14 essential topics and three specific to sports apps. Through a more detailed analysis of the reviews’ sentiment, we identified that users are generally happy about the applications. However, quality and advertisement are the most concerned factors for the users.
 %Our results pointed out the disproportionality of the popularity of sports to market but competitiveness amongst the functionality offered by the apps. Moreover, results showed a significant portion of apps using statistical and predictive analysis. We also concluded that limiting perspectives from global to specific helps to identify unique views. For example, analyzing more number of application in a specific category can unravel more details than general studies. Finally, analyzing the reviews’ sentiment will help narrow down issues faced by the users for different functionalities and sports types and can be recognized and mitigated in the early stage of app development.

%\balance{
%\bibliographystyle{ACM-Reference-Format}
\bibliographystyle{elsarticle-num}
\bibliography{sportsapps}

\end{document}